\documentclass[%
preprint,
nofootinbib,
 amsmath,amssymb,
 aps,
 11pt,
]{revtex4}

\usepackage{multirow}
\usepackage{booktabs}
\usepackage{cancel}
\usepackage{color}
\usepackage{graphicx,physics,slashed}
\usepackage{graphicx}
\usepackage{hyperref}
\hypersetup{colorlinks=true, citecolor=blue, urlcolor=blue, linkcolor=blue}
\usepackage{dcolumn}
\usepackage{bm}
\usepackage{subfigure}
\usepackage{slashed}
\usepackage{setspace} 
\usepackage{ulem}
\usepackage{soul} 
\usepackage{cancel} 

\begin{document}

\title{\Large Invisible and Semi-invisible Decays of Bottom Baryons}
\vspace{5mm}
 
\author{
Yong Zheng$^{1,2}$\footnote{yzheng2018@lzu.edu.cn}, 
Jian-Nan Ding$^3$\footnote{dingjn23@pku.edu.cn, corresponding author}, 
Dong-Hao Li$^{4,5}$\footnote{lidonghao@ihep.ac.cn, corresponding author},
Lei-Yi Li$^{4,5}$\footnote{lileiyi@ihep.ac.cn, corresponding author},
Cai-Dian L$\ddot{\rm u}$$^{4,5}$\footnote{lucd@ihep.ac.cn},
Fu-Sheng Yu$^{1,2,3}$\footnote{yufsh@lzu.edu.cn}
}

\address{
$^1$MOE Frontiers Science Center for Rare Isotopes, Lanzhou University, Lanzhou 730000, China\\
$^2$School of Nuclear Science and Technology, Lanzhou University, Lanzhou 730000, China \\
$^3$Center for High Energy Physics, Peking University, Beijing 100871, China \\
$^4$School of Physical Sciences, University of Chinese Academy of Sciences, Beijing 101408, China \\
$^5$Institute of High Energy Physics, CAS, P.O. Box 918(4) Beijing 100049, China
 }


\begin{abstract}
The similar densities of dark matter and baryons in the universe imply that they might arise from the same ultraviolet model. 
The B-Mesogenesis, which assumes dark matter is charged under the baryon number, attempts to simultaneously explain the origin of baryon asymmetry and dark matter in the universe. 
In particular, the B-Mesogenesis might induce  bottom-baryon decays into invisible or semi-invisible final states, which provide a distinctive signal for probing this scenario.
In this work, we systematically study the invisible decays of bottom baryons into dark matters, and semi-invisible decays of bottom baryons into a meson or a photon together with a dark matter particle. 
In particular, the fully invisible decay can explore the stable particles in B-Mesogenesis.
Some QCD-based frameworks are used to calculate the hadronic matrix elements under the B-Mesogenesis model. 
We estimate the constraints on the Wilson coefficients or the product of some new physics couplings with the Wilson coefficients by the semi-invisible and invisible decays of bottom baryons at future colliders.  
\end{abstract}


\maketitle
\vspace{10mm}

\section{Introduction}

Cosmology observations, such as the velocity dispersion of galaxies \cite{Zwicky:1933gu}, the Big Bang nucleosynthesis (BBN) \cite{Walker:1991ap,Cyburt:2015mya} and the cosmic microwave background (CMB) \cite{WMAP:2010qai}, provide strong evidence for the existence of dark matter (DM) and baryon asymmetry in the universe.
The cosmological measurements \cite{Planck:2018vyg} indicate the relic densities of dark matter ($\Omega_{\text{DM}}$) and baryons ($\Omega_\text{B}$) are at the same order, i.e. $\Omega_{\text{DM}}= (5.36 \pm 0.06) \ \Omega_{\text{B}}$, which implies the dark matter and baryon asymmetry might stem from the same ultraviolet (UV) model \cite{Davoudiasl:2010am}.

In recent years, a model of B-Mesogenesis was proposed to simultaneously explain the origins of DM and baryon asymmetry by assuming the dark matter charged under baryon number \cite{Elor:2018twp}.
The Sakharov conditions are satisfied in the following ways.
A long-lived particle produces bottom mesons and their anti-mesons out of the thermal equilibrium in the early universe. 
Then the neutral bottom mesons undergo $B^0-\overline B^0$ and $B_s^0-\overline B_s^0$ oscillations which provide CP violation naturally.  
Finally, the bottom mesons decay into an ordinary baryon with the baryon number $B=+1$ and a dark fermion $\psi$ which is charged under baryon number with $B=-1$, so that the visible baryon number is violated but the total baryon number is conserved. 
Last but not least, to preserve the DM and baryon asymmetry, the dark fermion $\psi$ should decay into a dark Majorana fermion $\xi$ and a dark scalar baryon $\phi$.
In this way, both the DM relic abundance and the baryon asymmetry in the universe can be simultaneously explained in this model. 
What's more, most of the new particles in this scenario locate at GeV scale, which can be precisely examined at the current B-factories and hadron colliders or some experiments in the near future.

This model has attracted a lot of attentions in both theories and experiments during the past a few years \cite{Nelson:2019fln,Alonso-Alvarez:2019fym,Alonso-Alvarez:2021qfd,Elahi:2021jia,Alonso-Alvarez:2021oaj,Elor:2022jxy,Berger:2023ccd,Khodjamirian:2022vta,Boushmelev:2023huu,Shi:2023riy,Shi:2024uqs,Belle:2021gmc,BaBar:2023dtq,BaBar:2023rer,Elor:2020tkc}. 
From the proposal of B-Mesogenesis scenario, most of the theoretical studies concentrate on the semi-invisible decays of B mesons to explore this scenario. 
The BABAR and Belle experiments have measured the decay of bottom mesons into a baryon and a dark sector anti-baryon $\psi$, such as $B^0\to \Lambda^0\psi$ and $B^+\to p\psi$ \cite{Belle:2021gmc,BaBar:2023dtq,BaBar:2023rer}.
On the other hand, it is also allowed that the bottom baryons decay into a meson and a dark baryon, such as $\Lambda_b(\Xi_b)\to P\bar{\psi}$, with $P$ denoting a light pseudo-scalar $\pi,K$ meson, which has been mentioned in the proposal of B-Mesogenesis \cite{Elor:2018twp} and well estimated in a recent study \cite{Shi:2024uqs}. 
However, for bottom baryons, there are other two processes which have not ever been studied under the B-Mesogenesis, which are the fully invisible decay ($\Lambda_b^0\to \xi\bar{\phi}$) and the semi-invisible radiative decay ($\Lambda_b^0\to\gamma\bar{\psi}$) of $\Lambda_b^0$ baryon.
In particular, the fully invisible decay of $\Lambda_b^0$ yields a distinctive signal to explore the stable dark particles in B-Mesogenesis. 
All these processes can be explored at the future colliders with a high accuracy. 
For instance, the upgrade of Belle-II \cite{Krizan:2022uir} is expected to produce the bottom baryons pair at the threshold, while the LHCb can measure the signals with its vertices detector \cite{Rodriguez:2021urv}.
What's more, as the Circular Electron Positron Collider (CEPC) \cite{CEPCStudyGroup:2018ghi} and Future Circular Collider (FCC-ee) \cite{FCC:2018evy} will produce substantial bottom baryons, these rare decays of bottom baryons are expected to be explored at the future lepton colliders. 
Therefore, to precisely probe the B-Mesogenesis, the systematical analysis on the invisible and semi-invisible decay of bottom baryons are crucial and might shed light on the study of dark matter and baryon asymmetry in the universe.

Except for $\Lambda_b^0\to \xi\bar\phi$, the fully invisible decays are also very interesting for neutron and hyperon decays, which share the same type of interaction as $u_ad_bd_c\psi$ with $a,b,c$ representing the flavors of up- and down-type quarks. 
It was proposed to explain the puzzle of neutron lifetime between the bottle method and the beam method, by allowing neutron decaying into dark matter \cite{Fornal:2018eol}.
It is also allowed for the invisible decays of $\Lambda^0$ hyperon in the new physics models similar to the B-Mesogenesis and the ones for the neutron lifetime anomaly \cite{Alonso-Alvarez:2021oaj}.
The BESIII collaboration has measured it as $BR(\Lambda^0\to \text{invisible}) < 7.4 \times 10^{-5}$ \cite{BESIII:2021slv}.
The invisible decays of bottom baryons rely on the interaction with a bottom quark, which is beneficial to explore the B-Mesogenesis.

When searching for the new physics in the decays of hadrons,  hadronic matrix elements are inevitable and crucial for determining the observables, except for the interaction of UV models. 
Therefore, it is necessary to calculate the corresponding matrix elements to examine the new physics or constrain the parameters of a specific model in hadron decays. 
Since the matrix elements relate to the strong interaction, we have to utilize a perturbative or non-perturbative method to determine the hadronic matrix elements. 
In practice, the decays of $\Lambda_b(\Xi_b)\to P\bar{\psi}$ involve  bottom baryon to light-meson transition form factors, which have been  investigated in the light-cone sum rules (LCSR) method very recently \cite{Shi:2024uqs}.
Since the baryon-to-meson form factors are a new kind of physical quantities, it deserves more efforts of detailed discussions in the theoretical studies. 
Under the heavy quark limit, these matrix elements can also be estimated within the QCD factorization approach \cite{Beneke:1999br,Beneke:2000ry,Beneke:2000wa,Bell:2020qus}. 
For example, this kind of form factors also emerge in the baryon and lepton number violation decays of $\Lambda_b\to P\ell$ from a Leptoquark model, and  have been studied very recently  within the QCD factorization \cite{Li:2024htn}. 
Although the higher dimensional operators of B-Mesogenesis are different to that of Leptoquark model, the bottom baryon to light meson transition form factors can be similarly determined within the QCD factorization. 
Our results suggest that the transition form factors can be factorized into a convolution of the perturbatively calculable hard-scattering kernel and the non-perturbative distribution amplitudes without endpoint divergence \cite{Li:2024htn,Wang:2011uv}.
The matrix elements involved in $\Lambda_b^0\to\gamma\bar{\psi}$ can also be calculated within the same QCD factorization method, while that of $\Lambda_b^0\to\xi\bar{\phi}$ can be estimated directly rely on the matrix elements in the SM. 
With these matrix elements, we can calculate the branching ratios of invisible and semi-invisible decays of bottom baryons, and predict the potential of exploring B-Mesogenesis at the future measurements.

This manuscript is organized as follows. 
In Sec.~\ref{sec:introduction-to-invisible-decay}, we give a brief introduction to the B-Mesogenesis scenario.
In Sec.~\ref{sec:b`baryon-dark-decays}, the hadronic matrix elements in  the invisible and semi-invisible decay of $\Lambda_b$ and $\Xi_b$ baryons are calcualted within the QCD factorization approach. 
In Sec.~\ref{sec:constraint-for-coupling}, the sensitivities on the parameters of B-Mesogenesis are discussed for the invisible and semi-invisible decay of bottom baryons at future collider.
Finally, we conclude in Sec.~\ref{sec:conclusion}.
\section{Brief Introduction to B-Mesogenesis
}\label{sec:introduction-to-invisible-decay}

As the relic densities of DM and net baryon are similar, they might arise from a same UV completion. 
In practice, the B-Mesogenesis \cite{Elor:2018twp,Alonso-Alvarez:2021qfd,Alonso-Alvarez:2021oaj,Elahi:2021jia}, which assume dark matter is charged under the baryon number, attempts to simultaneously explain the existence of DM and baryon asymmetry in the Universe.
In particular, this scenario will induce the decays of bottom baryons into dark baryons, which significantly violate the baryon number conservation in the visible sector.
The corresponding Lagrangian is given by:
\begin{equation}
\begin{aligned}\label{equ:Full-Lagrangian}
    \mathcal{L}_{int} \supset
    &-y^R_{u_a d_b} \epsilon_{ijk} Y^{*}_i \bar{u}_{R a}^{j} d_{R b}^{k, \mathcal{C}}
   -y_{u_a d_b}^{L} \epsilon_{ijk} Y^*_{i} \bar{u}_{L a}^{j} d_{L b}^{k,  \mathcal{C}}  - y_{\psi d_c} Y_i \bar{\psi} d_{R c}^{i, \mathcal{C}} -y_d~\!\bar{\psi} \phi\xi + h.c..
    \end{aligned}
\end{equation}
The $u$ ($d$) represents the up-type (down-type) quarks. 
The indices $a,b,c$ ($i,j,k$) denote the quark flavors (colors), $L$ $(R)$ represents the left-handed (right-handed) component of quarks and $\mathcal{C}$ represents a charge conjugation.
$Y$ is a color-triplet scalar with electric charge $Q_Y=-1/3$, which couples to ordinary quarks in the first two terms, and to a right-handed quark and a dark fermion in the third term.
It is a TeV-scale particle which will be integrated out in the bottom hadron decays. 
$\psi$ is a dark fermion with the baryon number $B_\psi=-1$, which is singlet under the standard model gauge group so that can only couple to the right-handed down-type quarks under that color conservation. 
It can be found that the total baryon number is conserved in the interaction of $Y_i\bar\psi d_{R_c}^{i,\mathcal{C}}$, while the visible baryon number is violated.
This can be illustrated in FIG.~\ref{fig:model}. 
$\psi$ will completely decay into stable dark matters, a dark scalar $\phi$ with $B_{\phi}=-1$ and a dark Majorana fermion $\xi$.
Their quantum numbers are summarized in TABLE~\ref{tab:new`particle`quantum`number}.

%
 \begin{table}[]
 \begin{spacing}{2.0}
    \centering
    \caption{The quantum numbers of the new particles in B-Mesogenesis, including particle spin $S$, electric charge $Q$, and baryon number $B$.}\label{tab:new`particle`quantum`number}
    \begin{tabular}{c|ccccc}
    \hline\hline 
    \  Field  \qquad & \quad\;  $S$ \ \quad  & \quad \;$Q$ \quad \quad & \quad \;$B$ \quad \quad  \tabularnewline
\hline 
       $Y$  & $\quad0$ & $-\dfrac{1}{3}$ & $-\dfrac{2}{3}$  \\
       $\psi$  & $\quad\dfrac{1}{2}$  & $ 0$ & $-1$ \\
        $\xi$  & $\quad\dfrac{1}{2}$  & $ 0$ & $0$ \\
        $\phi$  & $\quad0$  & $ 0$ & $-1$ \\
\hline\hline
    \end{tabular}
    \end{spacing}
\end{table}

If the mass of color-triplet scalar $M_Y$ is much heavier than that of bottom baryons, we can integrate out such particles and obtain an effective Lagrangian related to the (semi-)invisible decay, which is given by
\begin{equation}{\label{Eq:effective-Lagragian1}}
    \mathcal{L}_{\text{EFT}}= C^{L,R}_{u_ad_b,d_c}~\!{\mathcal{O}}^{L,R}_{u_ad_b,d_c}+ C^{L,R^*}_{u_ad_b,d_c}~\!\bar{\mathcal{O}}^{L,R}_{u_ad_b,d_c} ,
\end{equation}
where $C^{L,R}_{u_ad_b,d_c}=y^{L,R^*}_{u_a d_b} y^*_{\psi d_c}/M_{Y}^2$ is the Wilson coefficients  and the effective operators $\mathcal{O}^{L,R}_{u_ad_b,d_c}$, $\bar{\mathcal{O}}^{L,R}_{u_ad_b,d_c}$ are given by
\begin{equation}\label{Eq:Effective-Operators}
    \begin{aligned}
    {\mathcal{O}}^{L}_{u_ad_b,d_c}=\epsilon_{ijk} \left( \bar{u}_{La}^{i , \mathcal{C}}~\!d_{Lb}^{j} \right)\left( \bar{\psi}^{\mathcal{C}}~\!d_{Rc}^{k} \right), &  \quad \bar{\mathcal{O}}^{L}_{u_ad_b,d_c}=\epsilon_{ijk} \left( \bar{u}_{La}^{i}~\!d_{Lb}^{j , \mathcal{C}} \right)\left( \bar{\psi}~\!d_{Rc}^{k,\mathcal{C}} \right), 
        \\
    {\mathcal{O}}^{R}_{u_ad_b,d_c}=\epsilon_{ijk}\left( \bar{u}_{Ra}^{i, \mathcal{C}}~\!d_{Rb}^{j} \right)\left( \bar{\psi}^{\mathcal{C}}~\!d_{Rc}^{k} \right) ,& 
        \quad \bar{\mathcal{O}}^{R}_{u_ad_b,d_c}=\epsilon_{ijk} \left( \bar{u}_{Ra}^{i}~\!d_{Rb}^{j , \mathcal{C}} \right)\left( \bar{\psi}~\!d_{Rc}^{k , \mathcal{C}} \right).
    \end{aligned}
\end{equation}
For convenience, 
we factorize out the external field $\psi$ from the effective operators and rewrite the Lagrangian as:
\begin{equation}{\label{Eq:effective-Lagragian2}}
    \begin{aligned}
        \mathcal{L}_{\text{EFT}}&=
  C^{L,R}_{u_ad_b,d_c}~\!\bar{\psi}^\mathcal{C}~\! {\mathcal{O}}^{'L,R}_{u_ad_b,d_c} + h.c.,
    \end{aligned}
\end{equation}
where the operators $\mathcal{O}^{'L,R}_{u_ad_b,d_c}$ are shown as
\begin{equation}\label{Eq:operator`p}  
    \mathcal{O}^{'L}_{u_ad_b,d_c}=\epsilon_{ijk}\ d_{Rc}^k \left( \bar{u}_{La}^{i,\mathcal{C} }~\!d_{Lb}^{j} \right), \quad
        \mathcal{O}^{'R}_{u_ad_b,d_c}=\epsilon_{ijk} \ d_{Rc}^k\left( \bar{u}_{Ra}^{i, \mathcal{C}}~\!d_{Rb}^j \right).
\end{equation} 
\section{Invisible and Semi-Invisible decays of bottom baryons}{\label{sec:b`baryon-dark-decays}}

\begin{figure}{}
      \centering
   ~~~\subfigure[]{\includegraphics[width=0.32\textwidth]{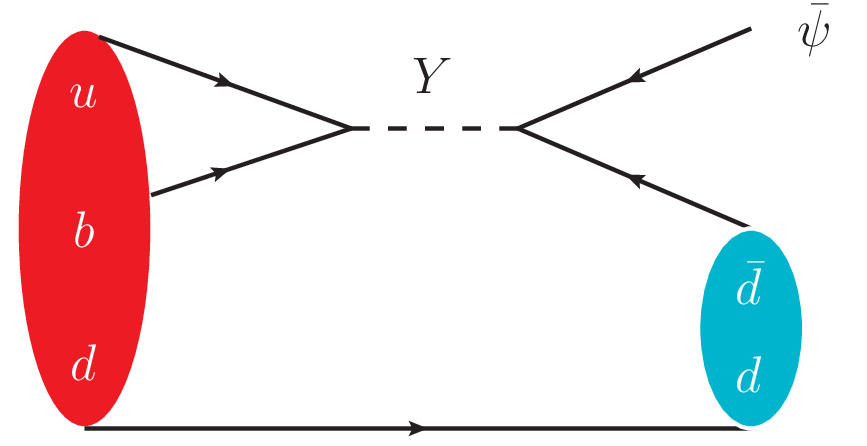}} 
    \subfigure[]{\includegraphics[width=0.3\textwidth]{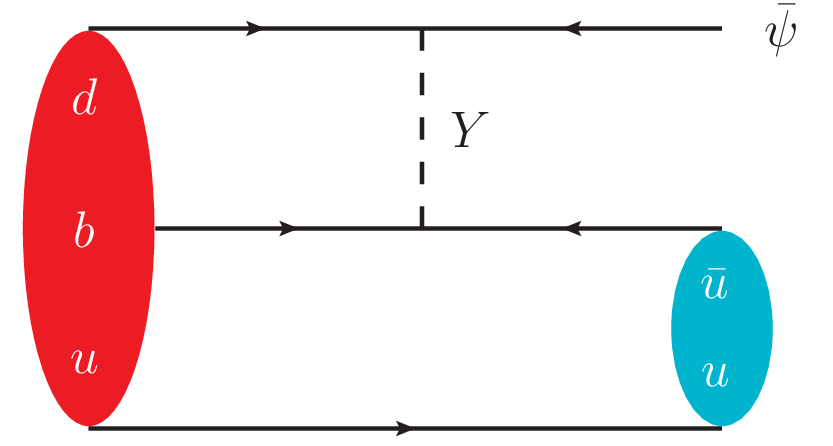}}
    \subfigure[]{\includegraphics[width=0.32\textwidth]{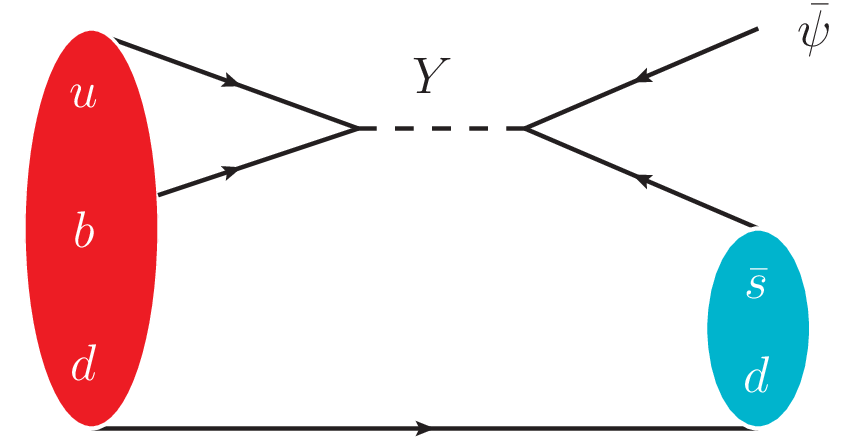}}

    ~~~\subfigure[]{\includegraphics[width=0.3\textwidth]{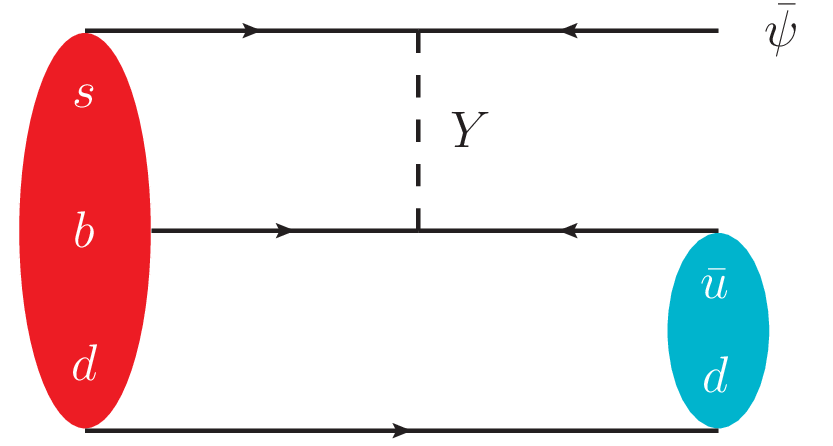}} 
    ~~\subfigure[]{\includegraphics[width=0.3\textwidth]{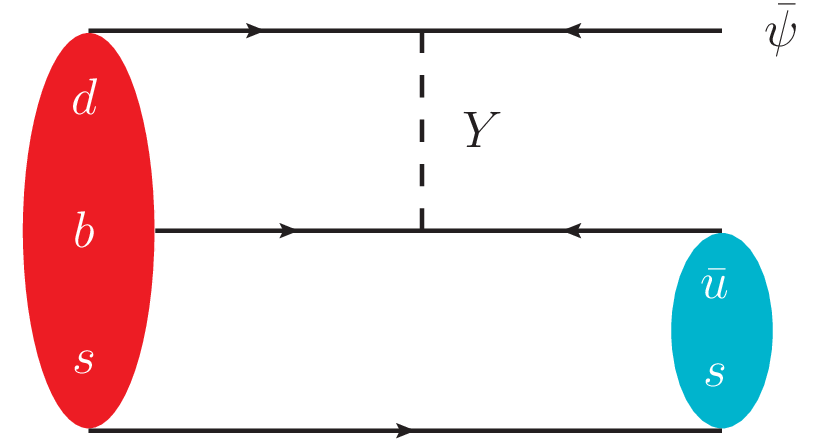}}
    ~~\subfigure[]{\includegraphics[width=0.3\textwidth]{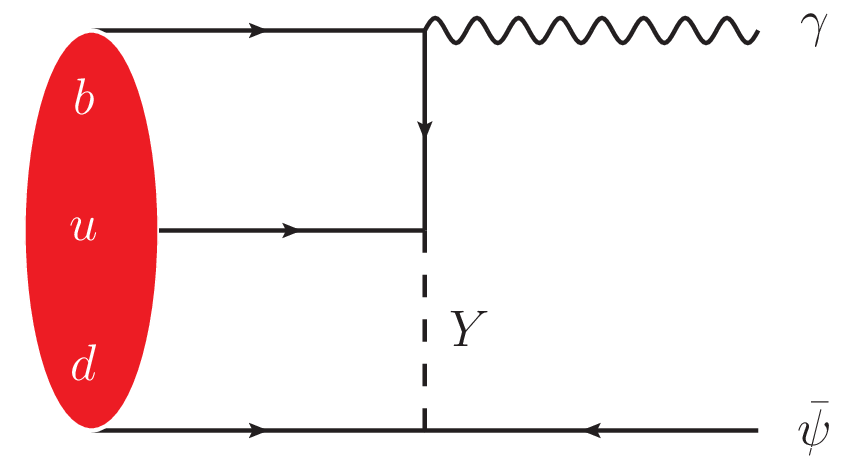}}

     ~~~~\subfigure[]{\includegraphics[width=0.3\textwidth]{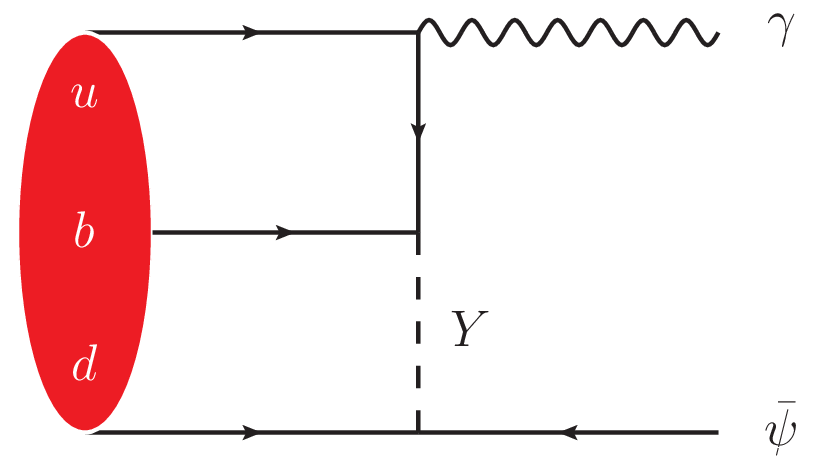}}     
    ~\subfigure[]{\includegraphics[width=0.3\textwidth]{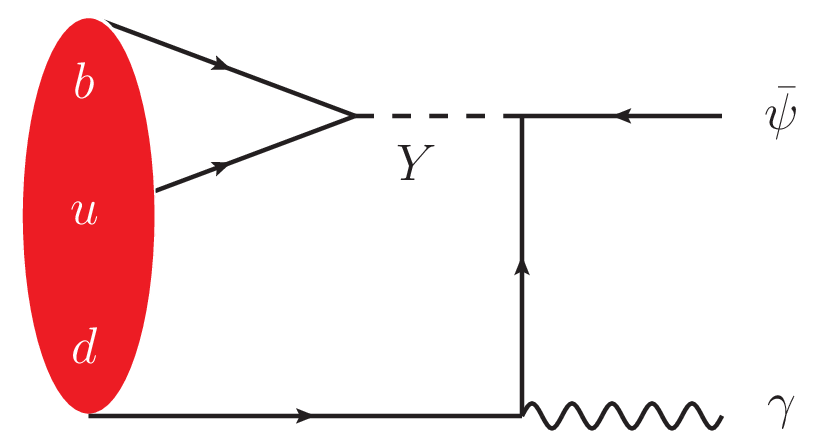}}
    ~~\subfigure[]{\includegraphics[width=0.3\textwidth]{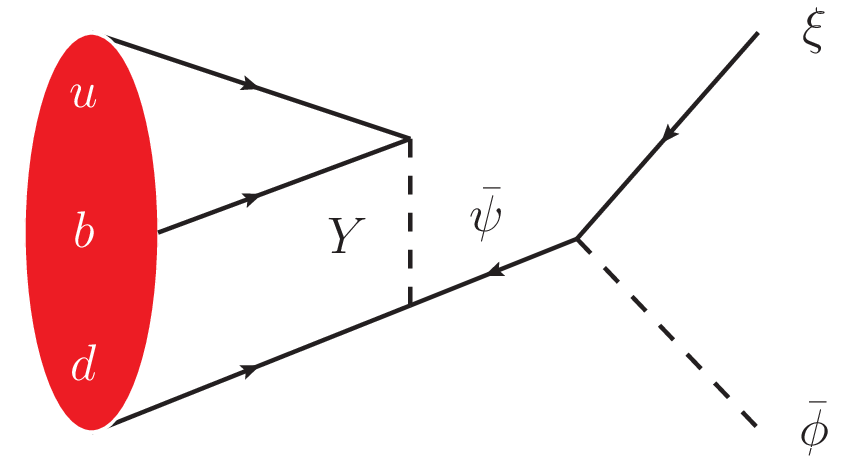}}
    
    \caption{The diagrammatic representations for the invisible and semi-invisible decays of bottom baryons in the B-Mesogenesis. (a) and (b) are the diagrams for the semi-invisible hadronic decay of $\Lambda_b^0\to \pi^0 \bar\psi$, (c) for $\Lambda_b^0\to K^0\bar\psi$, (d) for $\Xi_b^-\to \pi^-\bar\psi$, (e) for $\Xi_b^-\to K^-\bar\psi$, (f,g,h) for the semi-invisible radiative decay of $\Lambda_b^0\to \gamma\bar\psi$, and (i) for the fully invisible decay of  $\Lambda_b^0\to \xi\bar\phi$. The dark sector particles of $\bar\psi$ and $\bar \phi$ are charged under the baryon number of $B_{\bar\psi, \bar\phi}=+1$, so that the total baryon number are conserved in these processes. 
}
    \label{fig:model}
\end{figure}

With the effective Lagrangian of B-Mesogenesis in Eq.~\eqref{Eq:effective-Lagragian2}, we can systematically study the invisible and semi-invisible decays of bottom baryons.
To explore the B-Mesogenesis, we mainly focus on the exclusive decay modes of 
\begin{equation}
    \Lambda_b^0\to P^0\bar{\psi},~~\Xi_b^-\to P^-\bar{\psi},~~ \Lambda_b^0\to\gamma\bar{\psi},~~\Lambda_b^0\to\xi\bar{\phi} .
\end{equation}
The diagrammatic representation of these processes are shown in FIG.~\ref{fig:model}.
In the heavy quark limit, we calculate the relevant diagrams and apply the QCD factorization method for these heavy baryon to meson transition form factors at the leading order of $\alpha_s$.
The non-perturbative inputs are the heavy baryon and light meson distribution amplitudes, and we restrict our calculation to the leading twist (twist-2) distribution amplitudes.
Finally, the transition matrix elements defined by effective operators $\mathcal{O}^{'L,R}_{u_ad_b,d_c}$ can be represented as a convolution of the perturbative part with the leading twist heavy baryon and light meson distribution amplitudes.

\subsection{Hadronic semi-invisible decays of $\Lambda_b/\Xi_b\to P \bar{\psi}$ }
\begin{figure}
    \centering
        \includegraphics[width=0.9\textwidth]{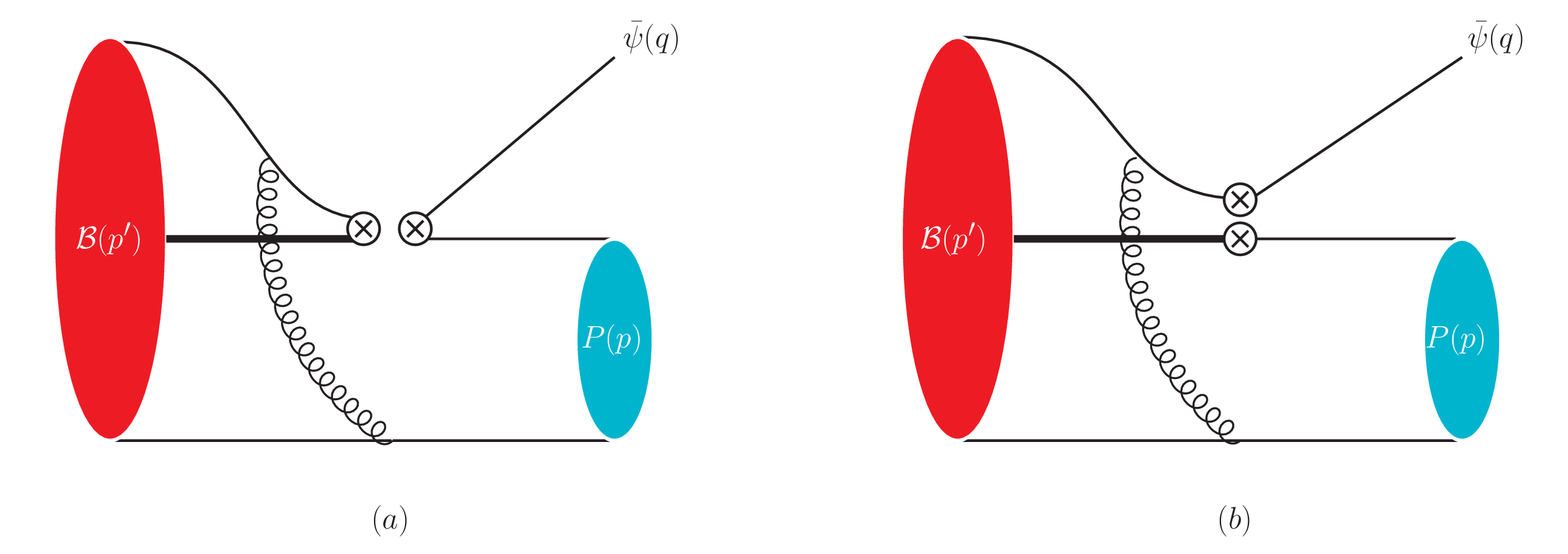}
\caption{ The Feynman diagrams of the $\Lambda_b^0/\Xi_b^- \to P\bar{\psi}$ decays, where the bold line represents the $b$ quark and the circle cross denotes the currents in the $\mathcal{O}^{L,R}$ operators.}
    \label{fig:hadron-final-states}
\end{figure}
In this subsection, we construct  the factorization formula and calculate the hadronic matrix elements of $\Lambda_b^0/\Xi_b^- \to P^{0,-}\bar{\psi}$   decays explicitly in the heavy quark limit.
The relevant diagrams are shown in  FIG.~\ref{fig:hadron-final-states}. Using the effective Lagrangian given in Eq.~\eqref{Eq:effective-Lagragian2}, one can express the decay amplitude for the operator $\mathcal{O}^{'L,R}_{u_ad_b,d_c}$ as,
\begin{equation}{\label{Eq:baryon-meson-amplitude}}
     i \mathcal{M} =
C^{L,R}_{u_ad_b,d_c}
\bigg(
\bar{v}_{\psi}^{\mathcal{C}}(q)
    \langle P(p)|
    \mathcal{O}^{'L,R}_{u_ad_b,d_c} |\mathcal{B}(p')\rangle \bigg),
\end{equation}
where $v_{\psi}$ is the spinor of dark baryon $\bar{\psi}$ with momentum $q = p'-p$, and $\mathcal{B}$ represents $\Lambda_b^0$ or $\Xi_b^-$ with momentum $p'=m_{\mathcal{B}}v$. 
The hadronic matrix elements could be parameterized in terms of two form factors respectively,
\begin{equation}{\label{Eq:B-P-form`factors-def}}
    \begin{aligned}
          \langle P(p)|
    \mathcal{O}^{'L,R}_{u_ad_b,d_c} |\mathcal{B}(p')\rangle =  P_R \bigg( 
    \zeta^{\mathcal{B} \to P(L,R)}_1(q^2) + \frac{\slashed{q}}{m_{\mathcal{B}}} \zeta^{\mathcal{B} \to P(L,R)}_2(q^2) \bigg) u_{\mathcal{B}}(p').
    \end{aligned}
\end{equation}
We will calculate the form factors working in the rest frame of the heavy baryon and choose two light-cone vectors along the fast-moving final state particles with the conventions of
\begin{equation}
n_\mu=(1,0,0,1),\quad \bar{n}_\mu=(1,0,0,-1),\quad v^\mu=(1,0,0,0).
\end{equation}
 Then the four-momentum of the final state particles can be written as
\begin{equation}
\begin{aligned}
 q &=\bigg(E_q+|\vec{q}|,\; E_q-|\vec{q}|,\; 0 \bigg),\quad
p=\bigg(E_p-|\vec{p}|,\; E_p+|\vec{p}|,\; 0 \bigg),\\
 E_q &=\frac{m_\mathcal{B}^2+m^2_{\psi}-m_P^2}{2m_\mathcal{B}},\qquad \quad \; \,  E_p=\frac{m_\mathcal{B}^2-m^2_{\psi}+m_P^2}{2m_\mathcal{B}},\\
|\vec{q}|&=|\vec{p}|=\frac{\sqrt{\left(m_{\mathcal{B}}^2 - (m_{\psi}+m_P)^2 \right)\left(m_{\mathcal{B}}^2 - (m_{\psi}-m_P)^2 \right)}}{2 m_{\mathcal{B}}},
\end{aligned}
\end{equation}
with
\begin{equation}
    p^\mu=(n\cdot p)\frac{\bar{n}^\mu}{2}+(\bar{n}\cdot p)\frac{n^\mu}{2}+p_{\perp}^\mu\equiv (n\cdot p,\bar{n}\cdot p,p_{\perp}) .
\end{equation}
When computing the Feynman diagrams in the heavy quark limit, we list the momenta involved as
\begin{equation}
\label{mom}
p'=(m_\mathcal{B},m_\mathcal{B},0),\quad q=(m_\mathcal{B},m_\mathcal{B}\lambda^2,0),\quad
p=(m_\mathcal{B}\lambda^2,m_\mathcal{B},0),\quad
k=(\Lambda_{\text{QCD}},\Lambda_{\text{QCD}},\Lambda_{\text{QCD}}),
\end{equation}
where $k$ corresponds to the momentum of the light spectator quark in the baryon. The parameter $\lambda\sim\frac{m_{\psi}}{m_{\mathcal{B}}}\sim\frac{m_P}{m_\mathcal{B}}\sim \frac{\Lambda_{\text{QCD}}}{m_\mathcal{B}}$, which is vanishing in the heavy quark limit. 

When considering only the leading twist distribution amplitudes and the tree-level QCD calculations, the contribution of operator $\mathcal{O}^{'R}_{u_ad_b, d_b}$ in Eq.~\eqref{Eq:effective-Lagragian2} to the transition matrix element always vanish, since there is an odd number of Dirac matrices between the two chiral projection matrices $P_R=\frac{1+\gamma_5}{2}$ within the operator $\mathcal{O}^{'R}_{u_ad_b, d_b}$.
For example, in the heavy quark limit, the Dirac structure of $\Lambda_b^0\to \pi^0\bar{\psi}$ transition amplitude in FIG.~\ref{fig:hadron-final-states}~(a) is given by
\begin{equation}
i\mathcal{M}\propto v_\psi^T \mathcal{C} P_R \left(\slashed{p} \gamma_5\right) \gamma_\mu
\left(\slashed{\bar{n}}\gamma_5 \mathcal{C} \right)
(\gamma^{\mu})^T x \slashed{p}^T
P_{L,R}
u_{\Lambda_b},
\label{example PR}
\end{equation}
where $\left(\slashed{p} \gamma_5\right)$ and $\left(\slashed{\bar{n}}\gamma_5 \mathcal{C} \right)$ are the light-cone projectors corresponding to the leading twist distribution amplitudes of $\Lambda_b^0$ and $\pi^0$, respectively~\cite{Mannel:2011xg,Beneke:2000wa}. 
The subscripts of $P_{L,R}$ respect to the insertion of operators $\mathcal{O}^{L}_{u_a d_b d_c}$ or $\mathcal{O}^{R}_{u_a d_b d_c}$. 
Since there are odd number of Dirac matrices between $P_R$ and $P_{L,R}$,  the contribution of right handed operators $\mathcal{O}^{R}_{u_a d_b d_c}$ will vanish. 
The contributions of the right-handed operators in other semi-invisible decay of bottom baryons will vanish in the same way.

In the heavy quark limit, the transition amplitude in Eq.~\eqref{Eq:baryon-meson-amplitude} also vanishes for operators  $\mathcal{O}^{L}_{ud,b}$ and $\mathcal{O}^{L}_{us,b}$ due to the parity conservation.
Take $\Lambda_b^0\to \pi^0 \bar{\psi}$ as an example, the partonic amplitude\footnote{The partonic amplitude defined as replacing $|\Lambda_b^0(p')\rangle$ and $\langle P(p)|$ by $|u(k_1) d(k_2) b(m_bv)\rangle$ and $\langle \bar{q} (xp) q(\bar{x}p)|$ \cite{Wang:2015vgv}.}
can be factorized as a product of two transition matrix elements in the spin space,
$i \mathcal{M}  
 \propto 
 \langle\bar{\psi}(q)|\,\bar{\psi}^{\mathcal{C}} b_R \, | b(m_bv)\rangle\,
 \langle \bar{q} (xp)q(\bar{x}p)| \,  \bar{u}^{\mathcal{C}}_L d_L \,|u(k_1) d(k_2)\rangle.$ Then we will find that there is a trace in the second matrix element in our calculation, which is always zero because there is an odd number of $\gamma$ matrices. 
 And we find that the $\gamma_5$ in the operator $\bar{u}^{\mathcal{C}}_L d_L$ cannot contribute to the trace, because there are only two independent Lorentz vectors in this process to contract with the 4-dimensional antisymmetric tensor $\epsilon_{\mu\nu\rho\sigma}$.
 Actually, the $J^P$ quantum number of quark pair $(ud)$ in $\Lambda_b^0$ baryon is $0^+$ in the leading twist approximation, and $\langle q (xp)\bar{q}(\bar{x}p)| \,  \bar{u}^{\mathcal{C}}_L d_L \,|u(k_1) d(k_2)\rangle$ describes a scalar decay into a pseudoscalar induced by a scalar operator, which violate parity conservation.
 The hadronic effects will not affect this conclusion cause the perturbative results in the final factorization formula are independent of the external state \cite{Descotes-Genon:2002crx}. 

In the following calculation, we only consider the contribution of the operator $\mathcal{O}^{'L}_{ub,d}$ or $\mathcal{O}^{'L}_{ub,s}$. 
Strange number conversed or violated processes are induced by corresponding operators, which is summarized as,
\begin{equation}
    \begin{aligned}
        \Lambda_b^0 \to \pi^0 \bar{\psi} \quad \text{and} \quad \Xi_b^- \to K^- \bar{\psi}, \qquad \text{induced by $\mathcal{O}^{'L}_{ub,d}$}, \\
         \Lambda_b^0 \to K^0 \bar{\psi} \quad \text{and} \quad \Xi_b^- \to \pi^- \bar{\psi}, \qquad \text{induced by $\mathcal{O}^{'L}_{ub,s}$}. \\
    \end{aligned}
    \label{operators}
\end{equation}

In the framework of QCD factorization approach, the longitudinal momentum of the quarks in the energetic final state meson is at the scale of $m_b$ and the exchanged gluon have virtuality of order $m_b\Lambda_{\text{QCD}}$, which can be treated perturbatively.
Then we calculate the hard-scattering contributions, such as the ones shown in FIG.~\ref{fig:hadron-final-states}.
The decay amplitudes are written as
\begin{equation}{\label{Eq:amplitudes`with`spinor}}
      i \mathcal{M} = \eta_{\Delta s}
C^{L}_{u_ad_b,d_c} \zeta^{\mathcal{B} \to P}
\bigg(
\bar{v}_{\psi}^{\mathcal{C}}(q) 
    P_{R}\frac{\slashed{n}}{2}\,
    u_{\mathcal{B}}(p{'}) \bigg),
\end{equation}
where the sign factor $\eta_{\Delta s}=-1$ for strange flavor conserved processes $\Lambda_b^0\to\pi^0$ and $\Xi_b^-\to K^-$, while $\eta_{\Delta s}=+1$ for strange flavor violated processes $\Lambda_b^0\to K^0$ and $\Xi_b^-\to \pi^-$.
The convention for the sign factor arises from the inconsistency between the quark flavor order in the operator $\mathcal{O}$ and the definition of the light-cone distribution amplitudes.

The factorization formula of the form factor $\zeta^{\mathcal{B} \to P}$ is given by
\begin{equation}{\label{eq:from`factor`analytical}}
\begin{aligned}
\zeta^{\mathcal{B} \to P} & =f_{P}f_{\mathcal{B}}^{(2)}
\int_{0}^{\infty}d\omega\,\omega\int_{0}^{1}du\int_{0}^{1}dx
\mathcal{J}^{\mathcal{B} \to P }(x,u,\omega,\mu)
\phi_{P}(x)
\psi_{2}(u,\omega),
\end{aligned}
\end{equation}
where $f_{P(\mathcal{B})}$ is the decay constant of particle $P(\mathcal{B})$, $\mathcal{J}^{\mathcal{B} \to P }(x,u,\omega,\mu)$ is the corresponding hard-scattering kernel and $\phi_P(x)$ and $\psi_2(u,\omega)$ are the leading twist light-cone distribution amplitudes of the $P$ meson and $\mathcal{B}$ baryon \cite{Ball:2008fw,Bell:2013tfa,Beneke:2000ry,Grossman:2015cak},
\begin{equation}
    \left\langle P(p)|\,[\bar{q}(t\bar{n})]_{A}\,[t\bar{n},0]\,[q(0)]_{B}\,|0\right\rangle =\dfrac{if_{P}}{4}\,\bar{n}\cdot p\,\bigg[\,\dfrac{\slashed{n}}{2}\,\gamma_{5}\,\bigg]_{BA}\int_{0}^{1}dx\,e^{ixt\bar{n}\cdot p}\,\phi_{P}(x,\mu),
    \label{pi LCDA}
\end{equation}
with the momentum fraction $x$ of anti-quark $\bar{q}$ in the meson $P$, and
\begin{equation}
\begin{aligned} 
&\bra{0}[u_{i}(t_{1}\bar{n})]_{A}\,[0,t_{1}\bar{n}]\,[d_{j}(t_{2}\bar{n})]_{B}\,[0,t_{2}\bar{n}]\,[b_{k}(0)]_{C}\ket{\Lambda_{b}^0(p{'})}
\\
= &\,\dfrac{1}{6} \dfrac{\epsilon_{ijk}}{4}\,f_{\Lambda_{b}}^{(2)}(\mu)\,[u_{\Lambda_{b}}(p{'})]_{C}\,\bigg[\,\dfrac{\slashed{\bar{n}}}{2}\,\gamma_{5}\,C^{T}\,\bigg]_{BA}\int_{0}^{\infty}d\omega\,\omega\int_{0}^{1}du\,e^{-i\omega(t_{1}u+t_{2}\bar{u})}\,\psi_{2}(u,\omega),  
\\
 &\bra{0}[d_{i}(t_{1}\bar{n})]_{A}\,[0,t_{1}\bar{n}]\,[s_{j}(t_{2}\bar{n})]_{B}\,[0,t_{2}\bar{n}]\,[b_{k}(0)]_{C}\ket{\Xi_{b}^{-}(p{'})}
 \\
= &\, \dfrac{1}{6}\dfrac{\epsilon_{ijk}}{4}\,f_{\Xi_{b}}^{(2)}(\mu)\,[u_{\Xi_{b}^{-}}(p{'})]_{C}\,\bigg[\,\dfrac{\slashed{\bar{n}}}{2}\,\gamma_{5}\,C^{T}\,\bigg]_{BA}\int_{0}^{\infty}d\omega\,\omega\int_{0}^{1}du\,e^{-i\omega(t_{1}u+ t_{2}\bar{u})}\,\psi_{2}(u,\omega),
\end{aligned}
\label{Baryon LCDA}
\end{equation}
where $\omega_i=(n\cdot k_i)$ represent the momentum components of the spectator quark respectively.
We use the convention $\omega_1=u \omega$, $\omega_2=\bar{u} \omega$ with $\omega=\omega_1+\omega_2$, where $u$ and $\bar{u}$ represent the momentum fraction of the spectator quark in the light quark pair. 
The coupling constants $f_{\mathcal{B}}$ for $\Lambda_b^0$ and $\Xi_b^-$, in the
heavy-quark limit, are defined by \cite{Bell:2013tfa}, respectively,
\begin{equation}{\label{Eq:definition-decay-constants}}
\begin{aligned}
    \epsilon_{ijk} \langle 0| u^{iT}_{\alpha}(0) d^{j}_{\beta}(0) b^{k}_{\gamma}(0) |\Lambda_b^0(p{'})\rangle &\equiv \frac{1}{4} \left\{ f_{\Lambda_b}^{(1)} [\gamma_5 C^{T}]_{\beta\alpha} + f_{\Lambda_b}^{(2)} [\slashed{v}\gamma_5 C^{T}]_{\beta\alpha}\right\} \left[u_{\Lambda_b}(p{'})\right]_{\gamma},\\
    \epsilon_{ijk} \langle 0| d^{iT}_{\alpha}(0) s^{j}_{\beta}(0) b^{k}_{\gamma}(0) |\Xi_b^-(p{'})\rangle &\equiv \frac{1}{4} \left\{ f_{\Xi_b}^{(1)} [\gamma_5 C^{T}]_{\beta\alpha} + f_{\Xi_b}^{(2)} [\slashed{v}\gamma_5 C^{T}]_{\beta\alpha}\right\} \left[u_{\Xi_b}(p{'})\right]_{\gamma}.
\end{aligned}
\end{equation}

%
%
The gauge links in Eq.~(\ref{Baryon LCDA}) is used to preserve the gauge invariance,
\begin{equation}
[t\bar{n},0]=\mathrm{P}\,\mathrm{exp}\bigg[\,ig\int_{0}^{t}dx\,\bar{n}\cdot A(x\bar{n})\,\bigg] ,
\end{equation}
which do not affect our calculations at the leading order of QCD.
The leading-twist LCDAs of pseudoscalar mesons can be expanded using
Gegenbauer polynomials \cite{Braun:2003rp},
\begin{equation}
\phi_{P}(x,\mu)=6x(1-x)\bigg[\,1+\sum_{n=1}^{\infty}a_{n}^{P}(\mu)\,C_{n}^{(3/2)}(2x-1)\,\bigg],
\end{equation}
and the leading-twist wave function
for the $\Lambda_{b}^0$ baryon is given by \cite{Ball:2008fw,Bell:2013tfa,Wang:2015ndk},
\begin{equation}
\psi_{2}(u,\omega)=u\,(1-u)\,\omega^{2}\,\dfrac{1}{\omega_{0}^{4}}\,e^{-\omega/\omega_{0}}.
\end{equation}
%
In the limit of the SU(3) flavor symmetry, the wave function of $\Xi_b^-$ is  the same as that of $\Lambda_b^0$. 
TABLE~\ref{tab:Input_Para} summarizes the numerical input parameters.
\begin{table} 	
\caption{The input parameters with renormalization scale $\mu_{0}=1\,\mathrm{GeV}$.} 	
\begin{spacing}{1.8} 		
\noindent \centering{}%
\begin{tabular}{ll|ll} 			
\hline 			
\hline  			
$\quad m_{\Lambda_{b}}=5.6196\,\mathrm{GeV}$ & \cite{ParticleDataGroup:2022pth}	& $\quad\tau_{\Lambda_{b}}=1.471\,\mathrm{ps}$ & \cite{ParticleDataGroup:2022pth} \\ 
$\quad m_{\Xi_{b}}=5.7970\,\mathrm{GeV}$ & \cite{ParticleDataGroup:2022pth}	& $\quad\tau_{\Xi_{b}}=1.572\,\mathrm{ps}$ & \cite{ParticleDataGroup:2022pth} \\			%
$\quad\omega_{0}=0.280\,_{-0.038}^{+0.047}\,\mathrm{GeV}$ & \cite{Wang:2015ndk} $\quad$ & 	$\quad f_{\Lambda_{b}}^{(2)}(\mu_{0})=0.030\pm 0.005\,\mathrm{GeV}^{3}$ \quad & \cite{Wang:2015ndk}$\quad$\\ 			
$\quad f_{\pi}=0.1304\pm0.0002\,\mathrm{GeV}$ \quad& \cite{Grossman:2015cak} $\quad$&	$\quad f_{K}=0.1562\pm0.0007\,\mathrm{GeV}$ & \cite{Grossman:2015cak} \\ 		
$\quad a_{1}^{\pi}(\mu_{0})=0$ & \cite{Grossman:2015cak} &	$\quad a_{2}^{\pi}(\mu_{0})=0.29\pm0.08$ & \cite{Grossman:2015cak}\\ 
$\quad a_{1}^{K}(\mu_{0})=-0.07\pm0.04$ & \cite{Grossman:2015cak} $\quad$ &	$\quad a_{2}^{K}(\mu_{0})=0.24\pm0.08$ & \cite{Grossman:2015cak}\\ 
\hline  			
\hline 		
\end{tabular} 	
\end{spacing} 
\label{tab:Input_Para}
\end{table}

The corresponding hard-scattering functions are calculated as follows,
\begin{equation}
\begin{aligned}
    \mathcal{J}^{\Lambda_{b}^0\to\pi^{0}}(x,u,\omega,\mu)&=\frac{1}{\sqrt{2}}\big(\mathcal{J}_u(x,u,\omega,\mu)+\mathcal{J}_{\bar{u}}(x,u,\omega,\mu)\big),\\
    \mathcal{J}^{\Lambda_{b}^0\to K^{0}}(x,u,\omega,\mu)&=\mathcal{J}^{\Xi_{b}^{-}\to K^{-}}(x,u,\omega,\mu)=\mathcal{J}_{\bar{u}}(x,u,\omega,\mu),\\
    \mathcal{J}^{\Xi_{b}^{-}\to\pi^{-}}(x,u,\omega,\mu)&=\mathcal{J}_u(x,u,\omega,\mu),
\end{aligned}
\end{equation}
with
\begin{equation}
    \mathcal{J}_u(x,u,\omega,\mu)=\frac{1}{9}\frac{\pi\alpha_s(\mu)}{u \omega^2 x},\qquad
    \mathcal{J}_{\bar{u}}(x,u,\omega,\mu)=\frac{1}{9}\frac{\pi\alpha_s(\mu)}{\bar{u} \omega^2 x} ,
\end{equation}
where the $\frac{1}{\sqrt{2}}$ is isospin factor comes from $|\pi^0\rangle = \frac{1}{\sqrt{2}}(|u\bar{u}\,\rangle-|d\bar{d}\,\rangle)$.
Note that the  asymptotic behaviors of the leading twist light-cone distribution amplitudes in the endpoint region are
\begin{equation}
    \phi_P(x)\sim x(1-x),\qquad  \psi_2(u,\omega)\sim u(1-u)\omega^2,
\end{equation}
which will cancel the divergent behavior in hard-scattering function $\mathcal{J}_u(x,u,\omega,\mu)$ and $\mathcal{J}_{\bar{u}}(x,u,\omega,\mu)$. Then the convolution integral in Eq.~\eqref{eq:from`factor`analytical} converges as noticed already in \cite{Li:2024htn,Wang:2011uv}.
Next, we  use the QCD
coupling constants as $\alpha_{s}(\small{2~ \mathrm{GeV}})\approx 0.3$ and other inputs summarized in TABLE~\ref{tab:Input_Para}. The numerical
results for the form factors are
\begin{equation}
\begin{aligned}\label{equ:FormFactorMeson}
\zeta^{\Lambda_{b}^0\to\pi^{0}} & =(1.43\,_{-0.54}^{+0.46})\times10^{-2}\;[\mathrm{GeV}^2],\\
\zeta^{\Lambda_{b}^0\to K^{0}} & =(1.23\,_{-0.47}^{+0.40})\times10^{-2}\;[\mathrm{GeV}^2], \\
\zeta^{\Xi_{b}^{-}\to K^{-}} & =(1.23\,_{-0.47}^{+0.40})\times10^{-2}\;[\mathrm{GeV}^2],\\
\zeta^{\Xi_{b}^{-}\to\pi^{-}} & =(1.01\,_{-0.38}^{+0.33})\times10^{-2}\;[\mathrm{GeV}^2].
\end{aligned}
\end{equation}
Note that the mass dimension of form factor is $[\zeta]=2$, since the dimension of the baryon coupling constant $[f_{\mathcal{B}}]=3$.

The from factors in Eq.~(\ref{equ:FormFactorMeson}) are related to the form factors $\zeta_1$ and $\zeta_2$ from the matching between Eq.~(\ref{Eq:amplitudes`with`spinor}) and 
Eqs.~(\ref{Eq:baryon-meson-amplitude},\,\ref{Eq:B-P-form`factors-def}),
\begin{equation}
\label{relationship}
\eta_{\Delta s}\zeta^{\mathcal{B}\to P}=\zeta_1=-\zeta_2,
\end{equation}
Keep in mind that we take the heavy quark limit during the calculation for heavy baryon to light meson transition form factors, and the reliability of our results is best when the momentum transfer $q^2=0$. 
In most of the calculations of heavy to light decays, for example $B\to\pi \ell\bar{\nu}_{\ell}$ \cite{Wang:2015vgv} or $\Lambda_b\to\Lambda \ell\bar{\ell}$ \cite{Wang:2015ndk}, the relevant form factors can be determined only at one or two values of $q^2$ and assumed a $q^2$ dependence to obtain the numerical results at arbitrary $q^2$ . 
The parameterization and extrapolation of the transition form factors have been extensively studied, we introduce the simplest single-pole model to extrapolate form factors $\zeta_{1,2}(q^2=0)$ toward $\zeta_{1,2}(q^2=m_{\psi}^2)$ \cite{Wirbel:1985ji,Shi:2024uqs,Lellouch:1995yv,Blake:2022vfl},
\begin{equation}
\label{sig pol}
    \zeta_{1,2}(q^2)=\frac{1}{1-q^2/m^2_{\text{pole}}}\zeta_{1,2}(q^2=0).
\end{equation}
Here  $\zeta_{1,2}(q^2=0)$ are obtained through the values in Eq.~(\ref{equ:FormFactorMeson}) and the relationship between $\zeta^{\mathcal{B}\to P}$ and $\zeta_{1,2}$ in Eq.~(\ref{relationship}). 
The denominator is introduced to describe the threshold Behaviour below the onset of the continuum. 
Frankly, the form factor has a pole at the mass of the lowest state that related to the operators in Eq.~\eqref{operators}. 
For instance, $m_{\rm pole}$ is the mass of $\Lambda_b^0\,(\frac{1}{2}^+)$ for $\Lambda_{b}^0\to\pi^{0} \,/\Xi^-_{b}\to K^{-}$ transition, as $\Lambda_b^0\,(\frac{1}{2}^+)$ is the lowest state of operator $\mathcal{O}^{'L}_{ub,d}$, while $m_{\rm pole}$ is the mass of $\Xi_b^0\,(\frac{1}{2}^+)$ for $\Lambda_{b}^0\to K^{0}\,/\Xi^-_{b}\to \pi^{-}$ transition.

Starting from Eqs.~(\ref{Eq:baryon-meson-amplitude},\,\ref{Eq:B-P-form`factors-def}), the branching fractions of $\mathcal{B} \to P \bar{\psi}$ are expressed as  
\begin{equation}
\label{equ:BRmeson}
\begin{aligned}
\mathcal{BR}(\mathcal{B} \to P \bar{\psi})&=\frac{|\Vec{q}_{\psi}|}{16 \pi \Gamma_{\mathcal{B}} }\left|  C^L_{ub,d} \right|^2 \\
\times
\Bigg[
\zeta^2_1(q^2)+&\zeta^2_2(q^2)\frac{m_\psi^2}{m_{\mathcal{B}}^2}+
\left(\zeta^2_1(q^2)+\zeta^2_2(q^2)\frac{m_\psi^2}{m_{\mathcal{B}}^2} \right)\frac{m_\psi^2-m_P^2}{m_{\mathcal{B}}^2}
+4\zeta_1(q^2)\zeta_2(q^2)\frac{m_\psi^2}{m_{\mathcal{B}}^2} \Bigg],
\end{aligned} 
\end{equation}
where $(u_ad_b,d_c)$ represents $(ub,d)$ or $(ub,s)$ for effective operators $ \mathcal{O}^{L}_{ub,d}$ or $ \mathcal{O}^{L}_{ub,s}$, $|\Vec{q}_{\psi}|$ is the momentum of the dark baryon $\bar{\psi}$ 
\begin{equation}
   |\Vec{q}_{\psi}|= \frac{\sqrt{(m_{\mathcal{B}}^2 - (m_{\psi}+m_P)^2)(m_{\mathcal{B}}^2 - (m_{\psi}-m_P)^2)}}{2 m_{\mathcal{B}}},
\end{equation}
and $\Gamma_{\mathcal{B}}$ ($m_{\mathcal{B}}$) is the decay width (mass) of the baryon $\mathcal{B}$.

\subsection{Radiative semi-invisible decay of $\Lambda_b^0\to \gamma \bar\psi$}
\begin{figure}
\begin{centering}
\includegraphics[width=0.45\textwidth]{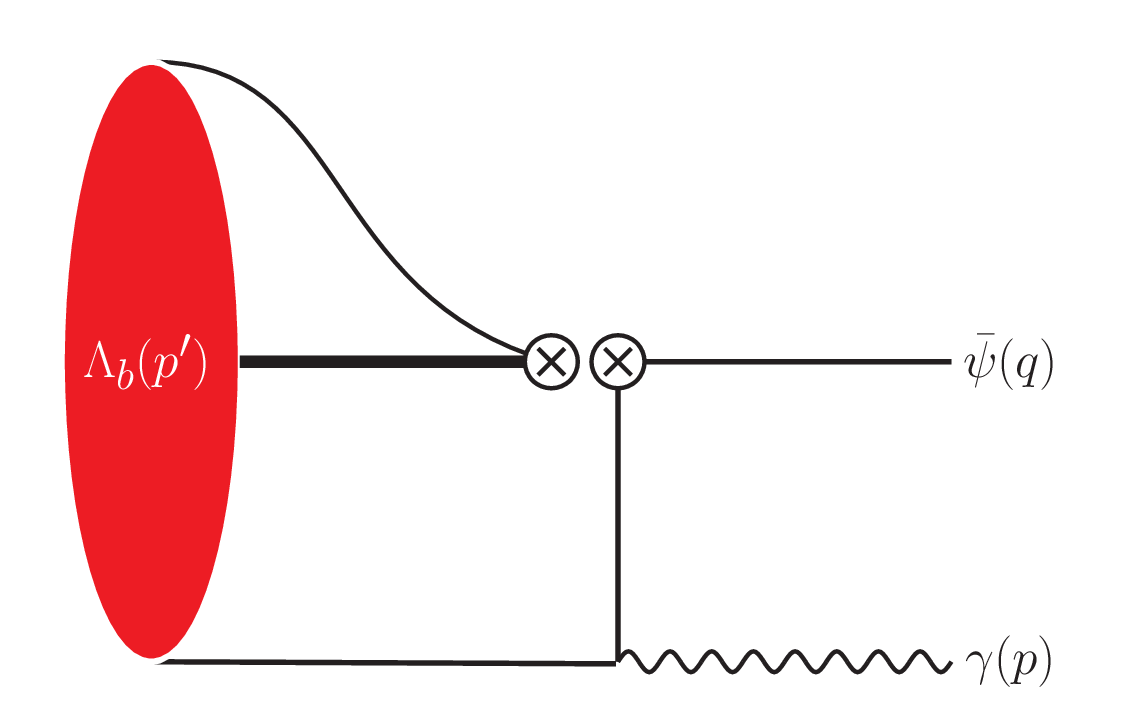}
\par\end{centering}
\caption{The Feynman diagram for $\Lambda_b^0 \to \gamma\bar{\psi}$ with the bold line representing the $b$ quark.}
\label{FIG:Lambda_b_to_gamma_psi}
\end{figure}
For the radiative decay processes $\Lambda_b^0 \to \gamma \bar{\psi}$ showed in FIG. \ref{FIG:Lambda_b_to_gamma_psi},
the decay amplitudes can be expressed as,
\begin{equation}{\label{Eq:baryon-gamma-amplitude}}
     i \mathcal{M} =
C^{L,R}_{u_ad_b,d_c}
\bigg(
\bar{v}_{\psi}^{\mathcal{C}}(q)
    \langle \gamma(p)|
    \mathcal{O}^{'L,R}_{u_ad_b,d_c} |\Lambda_{b}^0(p')\rangle \bigg).
\end{equation}
The transition amplitudes associated with the right-handed operators $\mathcal{O}_{ub,d}^{'R} $ and $ \mathcal{O}_{ud,b}^{'R}$ always vanish for the same reason we have mentioned previously in the hadronic semi-invisible decays.  
The hadronic matrix elements in Eq.~\eqref{Eq:baryon-gamma-amplitude} could be parameterized in terms of two form factors,
\begin{equation}\label{define gamma ff}
\langle \gamma(p)|
\mathcal{O}^{'L}_{u_ad_b,d_c} |\Lambda_{b}^0(p')\rangle 
= P_R \bigg( 
\zeta^{\Lambda_b \to \gamma}_1(q^2) \frac{\slashed{p}'}{m_{\Lambda_{b}}^2} i\sigma^{\mu\nu}  + \zeta^{\Lambda_b \to \gamma}_2(q^2) \frac{\slashed{q}}{m_{\Lambda_{b}}^2} i\sigma^{\mu\nu} 
\bigg) u_{\Lambda_b}(p')\epsilon^{*}_{\mu}p'_{\nu}.
\end{equation}

In the heavy quark limit, the transition amplitude for operator $\mathcal{O}^{'L}_{ud,b}$ vanishes in our calculation.
We can also arrive at this conclusion by considering the amplitude at the partonic level, which can be factorized as a product of two transition matrix elements in the spin space,
$i\mathcal{M}\propto  \langle\bar{\psi}(q)|\,\bar{\psi}^{\mathcal{C}} b_R \, | b(m_bv)\rangle\,
 \langle \gamma(p)| \,  \bar{u}^{\mathcal{C}}_L d_L \,|u(k_1) d(k_2)\rangle$.
The second matrix element describes the transition of a quark pair to the photon, with the initial momentum $k$,  the transfer momentum $r$ and the final state momentum $p$.
Its amplitude is proportional to the polarization vector $\epsilon_{\mu}$ of the photon.
Considering momentum conservation $k+r=p$ and on-shell condition $p^2=0$, there are two independent Lorentz-invariant amplitudes proportional to $p\cdot \epsilon$ and $r\cdot \epsilon$ respectively.
In the heavy quark limit, we neglect the momentum $k\sim$ $\Lambda_{\text{QCD}}$ of the spectator quarks. Therefore, the second transition matrix element must be proportional to $p \cdot \epsilon=r\cdot \epsilon$, which is zero after implementing the Ward Identity $p\cdot\epsilon=0$. 

The non-vanishing amplitude $\Lambda_b^0\to\gamma \bar{\psi}$ defined by $\mathcal{O}_{ub,d}^{'L}$ is given by,
\begin{equation}
\label{gamma heavy limit amplitude}
\begin{aligned}
i\mathcal{M}&=
C^{L}_{ub,d}  \bar{v}_{\psi}^{\mathcal{C}}(q)P_R\bigg(
\zeta^{\Lambda_{b}^0\to\gamma}_u
\dfrac{\slashed{\bar{n}}}{2}
\slashed{\epsilon}\dfrac{\slashed{n}}{2}
+
\zeta^{\Lambda_{b}^0\to\gamma}_d
\dfrac{\slashed{n}}{2}
\slashed{\epsilon}
\dfrac{\slashed{\bar{n}}}{2}
\bigg)
 u_{\Lambda_{b}}(p{'}),
\end{aligned}
\end{equation}
  The form factors can be factorized as
\begin{equation}
\begin{aligned}
\zeta^{\Lambda_{b}^0\to\gamma}_{u,d }
=f_{\Lambda_{b}}^{(2)}\int_{0}^{\infty}d\omega\,\omega\int_{0}^{1}du\mathcal{J}_{u,d}^{\Lambda_{b}^0\to\gamma}(u,\omega,\mu)\psi_{2}(u,\omega),
\end{aligned}
\end{equation}
where the index $u$\,($d$) denotes the contributions of photon emitting from $u$\,($d$) quark.
The photon radiation from $b$ quark is suppressed by $m_b$ in heavy quark propagator.
$\mathcal{J}_{u,d}^{\Lambda_{b}^0\to\gamma}(u,\omega,\mu)$ is the hard-scattering kernel for $\Lambda_b \to \gamma$ transition,
\begin{equation}
\mathcal{J}_u^{\Lambda_{b}^0\to\gamma}(u,\omega,\mu)=-\dfrac{1}{4}\bigg(\dfrac{Q_{u}(\mu)}{u\omega}\bigg),\qquad
\mathcal{J}_d^{\Lambda_{b}^0\to\gamma}(u,\omega,\mu)=-\dfrac{1}{4}\bigg(\dfrac{Q_{d}(\mu)}{\bar{u}\omega}\bigg),
\end{equation}
where the electric charge of the light quarks $u$ and $d$ are $Q_{u}(\mu)=\frac{2}{3}e(\mu)$ and $Q_{d}  (\mu)=-\frac{1}{3}e(\mu)$. 
Combining Eqs.~\eqref{Eq:baryon-gamma-amplitude}, \eqref{define gamma ff} and \eqref{gamma heavy limit amplitude}, we have
\begin{equation}
 \zeta_1^{\Lambda_{b}^0\to\gamma}=-\zeta_d^{\Lambda_{b}^0\to\gamma},\qquad \zeta_2^{\Lambda_{b}^0\to\gamma}=-\zeta_u^{\Lambda_{b}^0\to\gamma}+\zeta_d^{\Lambda_{b}^0\to\gamma}.
\end{equation}
The numerical result for the form factors of the spectator processes
$\Lambda_{b}^0\to\gamma\bar{\psi}$ is,
\begin{equation}
\begin{aligned}\label{equ:FormFactorGamma}
 \zeta^{\Lambda_{b}^0\to\gamma}_{u}=-(5.49\,_{-1.30}^{+1.18}) \times 10^{-3}\;[\mathrm{GeV^2}] ,\quad 
  \zeta^{\Lambda_{b}^0\to\gamma}_{d}=(2.74\,_{-0.65}^{+0.59}) \times 10^{-3}
  \;[\mathrm{GeV^2}],
\end{aligned}
\end{equation}
whose input is shown in TABLE~\ref{tab:Input_Para} with $\alpha_{em}\,(2\,\mathrm{GeV})\approx 1/133$ \cite{Erler:1998sy}. 
The single-pole model is also used to describe the $q^2$ dependence of form factor $\zeta_{1,2}(q^2)$,
\begin{equation}
\label{sig pol2}
    \zeta_{1,2}(q^2)=\frac{1}{1-q^2/m^2_{\Lambda_b}}\zeta_{1,2}(q^2=0),
\end{equation}
where $m_{\Lambda_b}$ is the mass of $\Lambda_b^0\,(\frac{1}{2}^+)$.
Then the branching ratio 
for $\Lambda_{b}^0\to\gamma\bar{\psi}$ with the operator $\mathcal{O}_{ub,d}^{'L}$ is  estimated as
\begin{equation}\label{equ:BRgamma}
    \begin{aligned}
    \mathcal{BR}(\Lambda_b^0 \to \gamma \bar{\psi})&
    =\frac{|\Vec{q}_{\psi}|}{8 \pi m^2_{\Lambda_b} \Gamma_{\Lambda_b} }\frac{1}{2}\sum_{\text{spin}}^{}\left| \mathcal{M} \right|^2\\
    &=\frac{|\Vec{q}_{\psi}|}{8 \pi  \Gamma_{\Lambda_b} }\left|  C^L_{ub,d} \right|^2 \left[\zeta_1^2(q^2)+
    \bigg(\zeta^2_1(q^2)+\zeta^2_2(q^2) \bigg) \frac{m_\psi^2}{m_{\Lambda_b}^2}+2\zeta_1(q^2)\zeta_2(q^2)\frac{m_\psi^2}{m_{\Lambda_b}^2}
    \right],
    \end{aligned}
\end{equation}
with $q^2=m_{\psi}^2$.
\subsection{Fully invisible decay for $\Lambda_b^0$}
\begin{figure}{}
        \centering
            \includegraphics[width=0.8\textwidth]{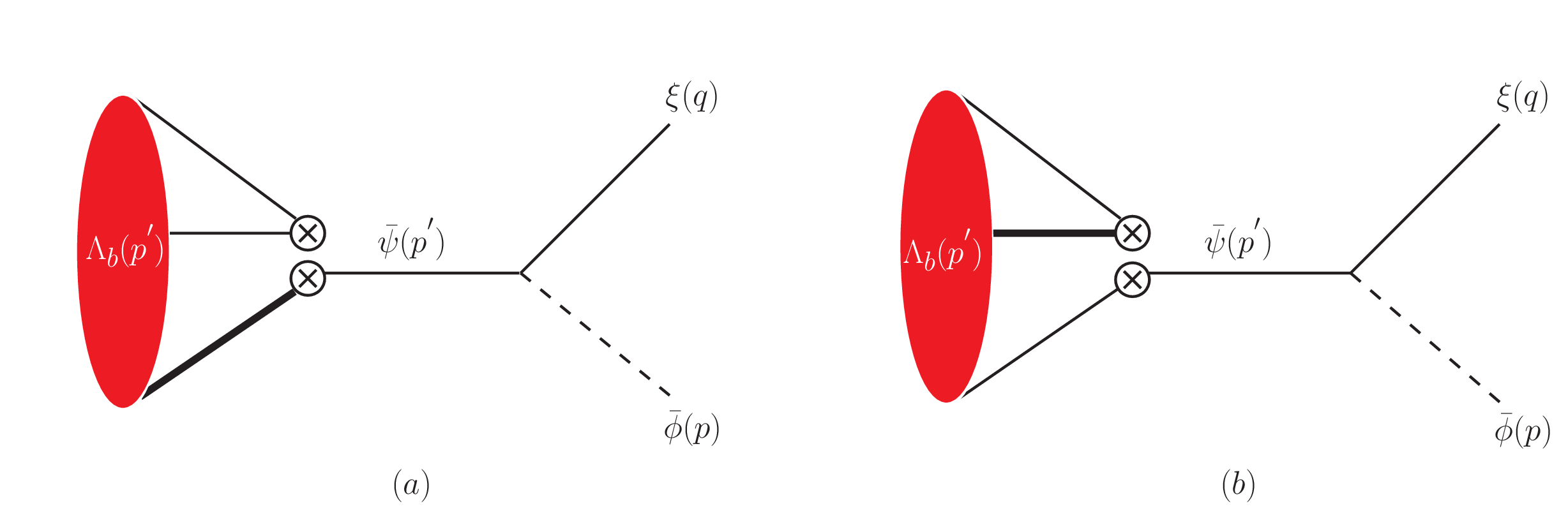}
    \caption{The Feynman diagram for $\Lambda_b^0 \to \xi\bar{\phi}$ process, with the bold line denoting the $b$ quark.}
    \label{FIG:Lambda_b_to_invisible}
\end{figure}
Under the B-Mesogenesis scenario, $\Lambda_b^0 \to \xi \bar{\phi}$ channel is governed by both effective operators ${\mathcal{O}}_{ud,b}^{'L,R}$ and ${\mathcal{O}}_{ub,d}^{'L,R}$.
The Feynman diagram is shown in FIG.~\ref{FIG:Lambda_b_to_invisible}.
The decay amplitudes could be respectively expressed as
\begin{equation}{\label{Eq:amplitudes`invisible`decay}}
    \begin{aligned}
        i \mathcal{M}_{ud,b}^{L,R} &=
    y_dC^{L,R}_{ud,b} \bigg(\bar{u}_{\xi}^{\mathcal{C}}(q)
    \frac{\slashed{p}{'}+m_{\psi}}{p'^{2}-m_{\psi}^2}  
  \langle  0| \mathcal{O}^{'L,R}_{ud,b} |\Lambda_b^0(p{'}) \rangle \bigg), \\
     i \mathcal{M}_{ub,d}^{L,R} &=
    y_dC^{L,R}_{ub,d} 
    \bigg(\bar{u}_{\xi}^{\mathcal{C}}(q)
    \frac{\slashed{p}{'}+m_{\psi}}{p'^{2}-m_{\psi}^2}  
    \langle 0|   \mathcal{O}^{'L,R}_{ub,d} |\Lambda_b^0(p{'})\rangle \bigg),
    \end{aligned}
\end{equation}
where $u_{\xi}(q)$ is the spinor for $\xi$ with the four-momentum $q$. 
To estimate the branching ratios of the invisible baryon decay, we have to firstly calculate the corresponding hadronic matrix elements in Eq.~\eqref{Eq:amplitudes`invisible`decay}, which could be parameterized in terms of the coupling constants defined by
\begin{equation}{\label{our_coupling}}
   \begin{aligned}
        \langle 0| \mathcal{O}^{'L,R}_{ud,b}  |\Lambda_b^0(p{'})\rangle = \lambda_{ud,b}^{L,R}\ P_R u_{\Lambda_b}(p{'}),
     \quad
     \langle 0| \mathcal{O}^{'L,R}_{ub,d} |\Lambda_b^0(p{'})\rangle = \lambda_{ub,d}^{L,R}\ P_R u_{\Lambda_b}(p{'}).
   \end{aligned}
\end{equation}
We found that the coupling $\lambda_{ud,b}^{L,R}$ and $\lambda_{ub,d}^{L,R}$ could be related to the coupling $f_{\Lambda_b}^{(1,2)}$ defined in Eq.~\eqref{Eq:definition-decay-constants} by multiply corresponding matrix element. For instance,
\begin{equation}
    \begin{aligned}
        \langle 0| \mathcal{O}^{'L,R}_{ud,b}  |\Lambda_b^0(p{'})\rangle  &=   \epsilon_{ijk} \langle 0|
      \left[u^{i T}\, C P_{L, R}\, d^j \right]b_{R}^k
      | \Lambda_b^0\left(p{'}\right) \rangle 
      \\
      &=\sum_{\rho} (CP_{L,R})^{\alpha\beta}(P_R)^{\rho\gamma} \left\{  \epsilon_{ijk} \langle 0| u^{iT}_{\alpha} d^{j}_{\beta} b^{k}_{\gamma} |\Lambda_b^0(p{'}) \rangle \right\}
      \\
      &= \frac{1}{4}\left\{ f^{(1)}_{\Lambda_b} \text{Tr}[CP_{L,R}\gamma_5C^T]+ f^{(2)}_{\Lambda_b}\text{Tr}[ CP_{L,R}\slashed{v}\gamma_5C^T ] \right\} P_R u_{\Lambda_b}(p{'}).
    \end{aligned}
\end{equation}
The last line is the result after using Eq.~\eqref{Eq:definition-decay-constants}. 
After the trace, we can obtain the relations,
\begin{equation}{\label{equ:InvisibleDecayConstant}}
    \begin{aligned}
        \lambda^L_{ud,b} &= -\frac{1}{2} f_{\Lambda_b}^{(1)}, \quad 
         \lambda^R_{ud,b} &= +\frac{1}{2} f_{\Lambda_b}^{(1)}.
    \end{aligned} 
\end{equation}
The same operation can be used for $\lambda^{L,R}_{ub,d}$, which can obtain,
\begin{equation}{\label{equ:InvisibleDecayConstant2}}
    \begin{aligned}
         \lambda^L_{ub,d} &= -\frac{1}{4} f_{\Lambda_b}^{(2)}, \quad
         \lambda^R_{ub,d} &= +\frac{1}{4} f_{\Lambda_b}^{(1)}.
    \end{aligned}
\end{equation}
For the numerical value of the couplings we quote the result of the NLO QCD sum rules
analysis in Ref.~\cite{Groote:1997yr}:
\begin{equation}
f_{\Lambda_b}^{(2)}\approx f_{\Lambda_b}^{(1)} = 0.030 \pm 0.005 \; [\text{GeV}^3].
\end{equation}
Then, we could estimate the branching ratios of the invisible baryon decay. 
Assuming the new physics individually exist in either the left-handed or right-handed component, the branching ratios of $\Lambda_b\to\xi\bar{\phi}$ is  given by
\begin{equation}\label{eq_branching-ratio}
    \begin{aligned}
    \mathcal{BR}({\Lambda_b\to\xi\bar{\phi}})^{L,R}_{u_ad_b,d_c}
    &=
    \frac{\left|\Vec{q}_{\xi}\right|}{8 \pi m_{\Lambda_b} \Gamma_{\Lambda_b}} 
    \bigg(\left|y_d\right|^2 \left|C^{L,R}_{u_ad_b,d_c}\right|^2
    \left|\lambda_{_{u_ad_b,d_c}}^{L,R}\right|^2 \bigg)
    \frac{E_{\xi}\left( m_{\Lambda_b}^2 + m^2_{\psi} \right) + 2 m_{\Lambda_b} m_{\psi}m_{\xi}}{\left(m_{\Lambda_b  }^2 -m_{\psi}^2\right)^2},
    \end{aligned}
\end{equation}
where $(u_ad_b,d_c)$ represents $(ud,b)$ or $(ub,d)$ for different effective operators $ \mathcal{O}^{L,R}_{ud,b}$ and $ \mathcal{O}^{L,R}_{ub,d}$ convention in Eq.~\eqref{Eq:Effective-Operators} respectively, $m_i~\!(i=\Lambda_b^0,\bar{\psi},\bar{\phi},\xi)$ and $\Gamma_{\Lambda_b}$ are the mass of particle $i$ and the total width of heavy baryons. 
The $\Vec{q}_{\xi}$ and $E_{\xi} = \frac{m_{\Lambda_b}^2  + m_{\xi}^2 - m_{\phi}^2}{2m_{\Lambda_b}}$ are the momentum and kinetic energy of $\xi$ in heavy baryon $\Lambda_b^0$ rest frame. 

\section{Probe B-Mesogenesis with invisible and semi-invisible decay of bottom baryons}
\label{sec:constraint-for-coupling}

With the hadronic matrix elements obtained in the above section, we can explore the B -Mesogenesis via the  invisible and semi-invisible decays of bottom baryons at the future lepton colliders. 
The reconstructions of the invisible and semi-invisible decays of bottom baryons are always difficult in experiments.
But it can be overcome by the double tag method \cite{MARK-III:1989dea}. 
In the $e^+e^-$ collisions, the initial energy is well known in each event.  
Suppose a good resolution in the detectors, we reconstruct all the recoiling particles except for $\Lambda_b^0/\Xi_b^{0,-}$ in the colliding event, which helps to determine the invariant mass of the bottom-baryon candidate using the energy-momentum conservation. 
It would be identified as a production of bottom baryon, if the invariant mass is consistent with the mass of $\Lambda_b^0/\Xi_b^{0,-}$ and the conservation laws are satisfied for the electric charge, the baryon number, the bottom number and the strange number. 
Then we are able to measure the invisible and semi-invisible decays of bottom baryons. 
The double tag method has been widely used by BESIII, BABAR and Belle. For example, BESIII measured the absolute branching fraction of $\Lambda_c^+\to \Lambda^0e^+\nu_e$ at the $\Lambda_c^+\bar\Lambda_c^-$ threshold, which determines $\Lambda_c^+$ by reconstructing the charmed anti-baryon via its hadronic decays such as $\bar\Lambda_c^-\to \bar p K^+\pi^-$ \cite{BESIII:2015ysy}. 
It is also possible to use the double tag method at the energy much higher than the baryon-antibaryon threshold.  
The Belle collaboration measured the absolute branching fractions of $\Lambda_c^+$ decays with reconstructing the recoiling $D^{(*)-}\bar p\pi^+$ system in the event of  $e^+e^-\to D^{(*)-}\bar p\pi^+\Lambda_c^+$ at the colliding energy around 10 GeV \cite{Belle:2013jfq}. 
Therefore, the invisible and semi-invisible decays of $\Lambda_b^0/\Xi_b^-$ can be measured using the double tag method at the $e^+e^-$ colliders, such as BelleII at the bottom baryon-antibaryon threshold, or CEPC and FCC-ee at a higher energy as well. 

In practice to estimate the sensitivity on the parameters of B-Mesogenesis, we define the significance $s$ of the signal at $95\%$ C.L. as
\begin{equation}
s=\frac{n_S}{\sqrt{n_S+n_B}}
=\sqrt{n_S}=\sqrt{N_{\mathcal{B}}\times Br(\mathcal{B}\to X)}
=2,
\end{equation}
where $n_S$, $n_B$ denote the events number of signal and background at future lepton colliders. 
The background events are negligible using the double tag method, since the three kinds of processes studied in this work are all forbidden in the Standard Model by the visible baryon number violation.  
The branching fractions  $Br(\mathcal{B}\to X)$ with $X=\pi\bar{\psi}, K\bar{\psi}, \gamma\bar{\psi}, \xi\bar{\phi}$ are given in Eqs.~(\ref{equ:BRmeson},\,\ref{equ:BRgamma},\,\ref{eq_branching-ratio}), respectively.
$N_\mathcal{B}$ represents the production number of bottom baryons detected by the double tag method.
Notice its difference from the total production of bottom baryons in the collisions.
Besides, we assume an efficiency of  reconstruction as $100\%$  for pion, kaon and photon in the semi-invisible decays. 
In this work, we assume the production number of bottom baryons by the double tag method are $N_\mathcal{B}=N_{\Lambda_b,\Xi_b}=10^8$ as a benchmark for the future lepton colliders. 
The corresponding constraints can directly scale up to the results respected to the actual production number of bottom baryons at future lepton colliders, which are still indistinct in the current theoretical study.

There are some constraints on the masses of the dark particles under the B-Mesogenesis scenario. 
To prevent the decay of proton into the dark particles, the masses of dark matters 
must satisfy
\begin{equation}
    \begin{aligned}
        m_{\bar{\psi}} &> m_p -m_e-m_{\nu_e}  \simeq 937.8\; \text{MeV},\\
        m_{\xi}+m_{\bar{\phi}} &> m_p -m_e-m_{\nu_e}  \simeq 937.8\; \text{MeV},
    \end{aligned}
\end{equation}
where the former prohibits $p\to \bar{\psi}e^+\nu_e$ decay, while the latter prevents $p \to \bar{\phi}\xi e^+\nu_e$ decay through an off-shell $\bar{\psi}$ propagation.
In addition, the stability of dark matter requires the mass difference between the dark scalar $\bar{\phi}$ and dark Majorana fermion $\xi$ must obey \cite{Alonso-Alvarez:2021qfd}
\begin{equation}{\label{equ_mass_splitting}}
    |m_{\xi}- m_{\bar{\phi}}| < m_p + m_e + m_{\nu_e} \simeq 938.8\ \text{MeV}.
\end{equation}
Otherwise, the dark particles could decay into each other by emitting a proton, an electron and a neutrino, which will diminish the generated baryon asymmetry in the Universe.

\subsection{Probe B-Mesogenesis in semi-invisible decays of bottom baryons.}

\begin{figure}{}
    \centering
    \!\!\subfigure[]{\includegraphics[width=0.48\textwidth]{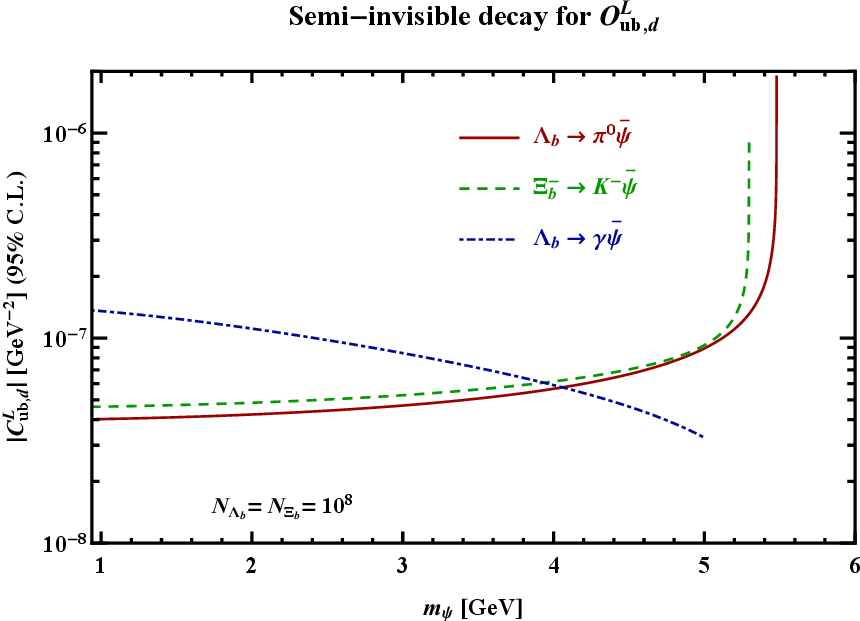}} 
    \subfigure[]{\includegraphics[width=0.48\textwidth]{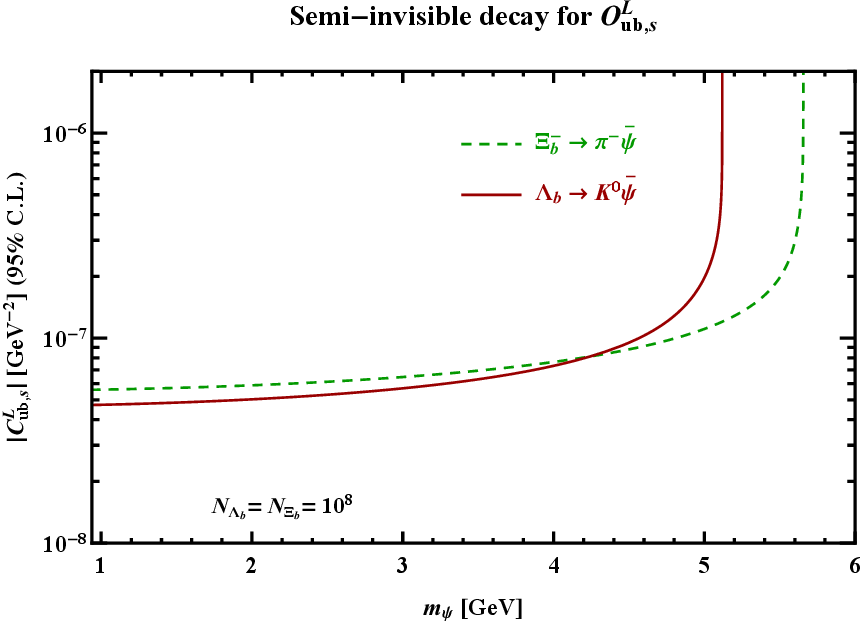}}
    \caption{The constraints on the Wilson coefficients at $95\%$ C.L. versus $m_\psi$ in the semi-invisible decays of bottom baryons.
    The left panel illustrates the constraints on $|C^L_{ub,d}|$, while the right panel display the constraints on $|C^L_{ub,s}|$.
    The declines of the constraints at the tail result from the suppression of phase space.
    When dark fermion mass $m_\psi$ close to the mass of $\Lambda_b^0$ baryon, the soft photon radiation in $\Lambda_b^0\to\gamma\bar{\psi}$ channel will induce a serious infrared divergence, which invalid the leading order estimation in this region.
    Hence, we omit the constraints in $\Lambda_b^0\to\gamma\bar{\psi}$ process when $m_\psi>5$ GeV (blue dot dashed line).
}
    \label{fig:constraint-Wilson-coefficient}
\end{figure}

Firstly, we estimate the sensitivity on the B-Mesogenesis by searching for the semi-invisible decay of bottom baryons at future lepton colliders.
Their branching ratios are only sensitive to the mass $m_\psi$ and Wilson coefficients $|C^L_{ub,d}|$ or $|C^L_{ub,s}|$ of the scenario. 
The constraints on the Wilson coefficients $|C^L_{ub,d}|$ and $|C^L_{ub,s}|$ versus $m_\psi$ at $95\%$ C.L. are illustrated in FIG.~\ref{fig:constraint-Wilson-coefficient}, where the production number of bottom baryons by the double tag method are assuming to be $N_{\Lambda_b,\Xi_b}=10^8$. 
The constraints on the right-handed operators $\mathcal{O}^R_{ud_a,d_b}$ are omitted as they could not project out the leading-twist wave function of $\Lambda_b$ and $\Xi_b$ baryons. 

For $\Lambda_b(\Xi_b)\to P\bar{\psi}$ processes, we found that the semi-invisible decay $\Lambda_b^0\to\pi^0\bar{\psi}$ is most sensitive to operator $\mathcal{O}^L_{ub,d}$, while the $\Lambda_b\to K^0\bar{\psi}$ 
process is most sensitive to operator $\mathcal{O}^L_{ub,s}$. 
When assuming the production number of bottom baryons by the double tag method is $N_{\Lambda_b,\Xi_b}=10^8$, the constraints on the Wilson coefficient can reach $\mathcal{O}(10^{-8})$ GeV$^{-2}$.  
The declines of the Wilson coefficients at the tail result from the suppression of phase space. 
Note that the form factors and the hadrons mass in semi-invisible decay are similar, the constraints on the Wilson coefficients are adjacent in these processes. 

For the $\Lambda_b^0\to\gamma\bar{\psi}$ channel, the constraints on the Wilson coefficients can reach at least $\mathcal{O}(10^{-7})$ GeV$^{-2}$.
The constraint is looser than that in the hadronic semi-invisible decay in small mass region, 
due to the form factors of photon radiation is smaller than those of meson final states. 
However, in the large mass region, the constraints will be tighter by the evolution of form factors.
What's more, when the dark fermion mass $m_\psi$ close to the mass of initial baryons $m_{\Lambda_b}$, the divergence in the evolution of form factors invalid the estimation of branching ratio. 
Therefore, we omit the constraints in $\Lambda_b^0\to\gamma\bar{\psi}$ channel when $m_\psi>5\;\text{GeV}$.

Last but not least, since the final state hadrons in $\Xi_b^-$ semi-invisible decay are all electric charged, these decay modes could be probed at higher energy future lepton colliders (i.e. CEPC or FCC-ee) with a displaced vertex method. 
Therefore, we expect a well exploration on these semi-invisible decay of bottom baryons at future lepton colliders.

\subsection{Probe B-Mesogenesis in invisible decay of bottom baryons.}

As shown in Ref. \cite{Elor:2018twp,Elor:2022jxy,Khodjamirian:2022vta,Boushmelev:2023huu,Shi:2023riy,Shi:2024uqs}, previous studies mainly focus on searching for the dark fermion $\psi$ in the semi-invisible decay of bottom baryons or mesons. 
To keep the dark matter to be stable, the B-Mesogenesis must involve a dark Majorana fermion $\xi$ and a dark scalar baryon $\phi$ and requires all the dark fermion $\psi$ completely decay into these two invisible particles, with $\mathcal{BR}(\psi\to \phi\xi)=100\%$.
Therefore, the coupling of $y_d$ and the interaction of $\bar\psi\phi\xi$ are very hard to be explored in the semi-invisible decays where $\psi$ is always on shell. 
The width of $\psi$ depending on $y_d$ cannot be measured in experiments since the final states $\phi$ and $\xi$ are not measurable.
On the contrary, the $\bar\psi$ is off-shell in the fully invisible decay of $\Lambda_b^0\to\xi\bar{\phi}$, seen in FIG.~\ref{fig:model}~(i) or FIG.~\ref{FIG:Lambda_b_to_invisible}. Then it is able to explore the stable dark matters $\xi$ or $\bar\phi$, and measure the interaction of $y_d\bar\psi\phi\xi$ as shown in the branching fraction of $\Lambda_b^0\to\xi\bar{\phi}$ in Eq.~\eqref{eq_branching-ratio}.
So we suggest that the fully invisible decay of $\Lambda_b^0\to\xi\bar{\phi}$ is crucial for probing the B-Mesogeneisis as it provides a distinctive signal to examine the stable dark particles in this scenario.

\begin{figure}
    \centering
    \subfigure[]{
\includegraphics[width=0.46\textwidth]{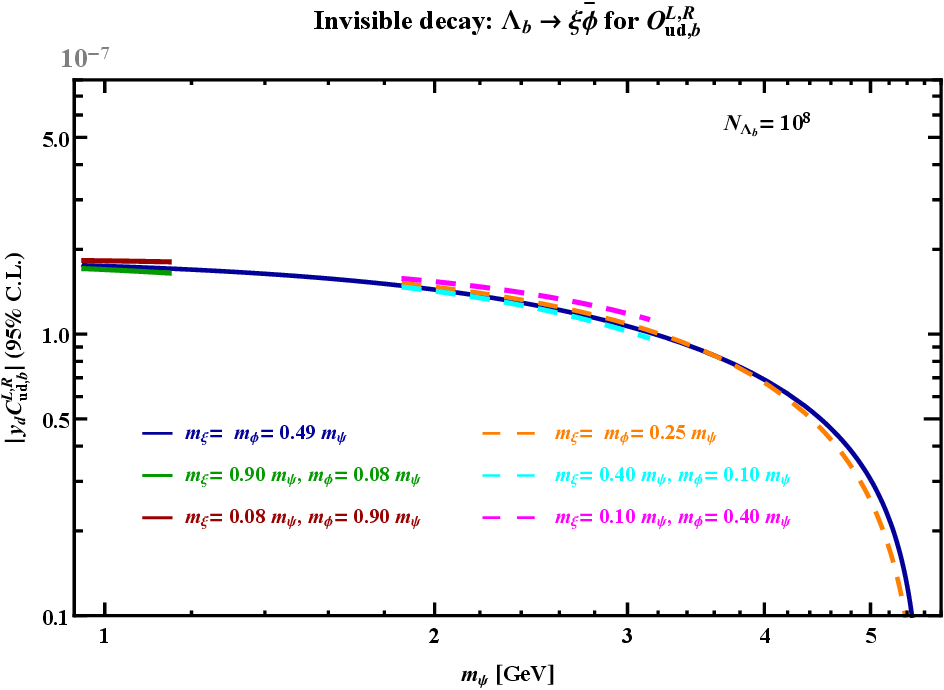}}
    ~~\subfigure[]{\includegraphics[width=0.46
\textwidth]{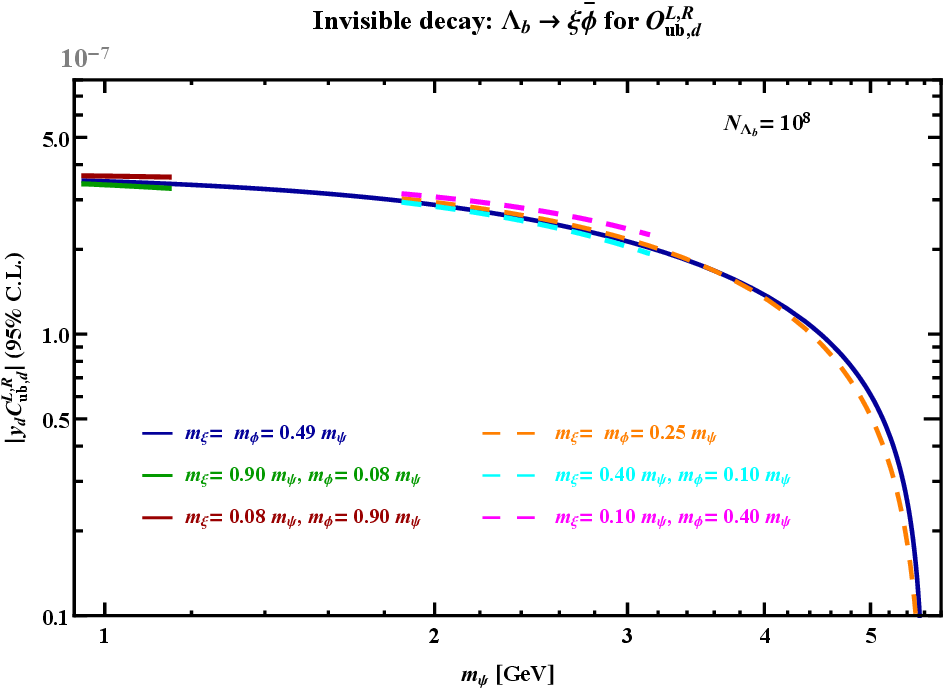}} 
\caption{The constraints on the parameters $|y_d~\!C^{L,R}_{ud,b}|$ and $|y_d~\!C^{L,R}_{ub,d}|$ versus $m_{\psi}$ from the invisible decay of $\Lambda_b^0$ with $95\%$ C.L. 
We estimate the constraints in six different mass distribution of dark matters.
The left and right panels illustrate the constraints related to operators $\mathcal{O}_{ud,b}^{L,R}$ and $\mathcal{O}_{ub,d}^{L,R}$, respectively.  
The dark matter (DM) stability in Eq.~\eqref{equ_mass_splitting} prohibits a large mass difference between dark fermion $\xi$ and dark scalar $\bar{\phi}$, which imposes a truncation at large mass region if $m_\phi\neq m_\xi$.
For the three cases where $m_\xi+m_\phi=0.5 \,m_\psi$ (dashed line), there is another truncation at small $m_\psi$ region to forbid the decay of proton into $\xi\bar{\phi}$ through an off-shell $\bar{\psi}$ particle.
}
    \label{fig:constraint-mchi}
\end{figure}

\begin{figure}{}
    \centering
    \!\!\subfigure[]{\includegraphics[width=0.45\textwidth]{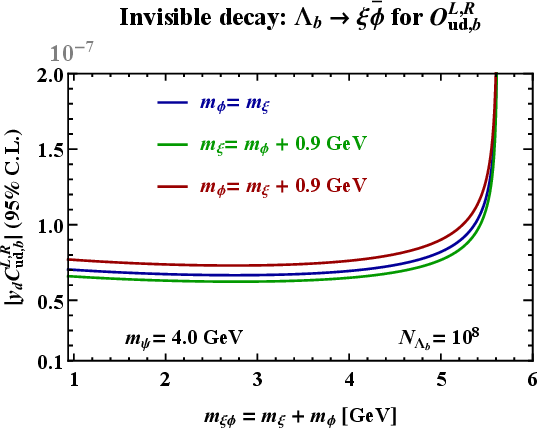}} 
    \subfigure[]{\includegraphics[width=0.45\textwidth]{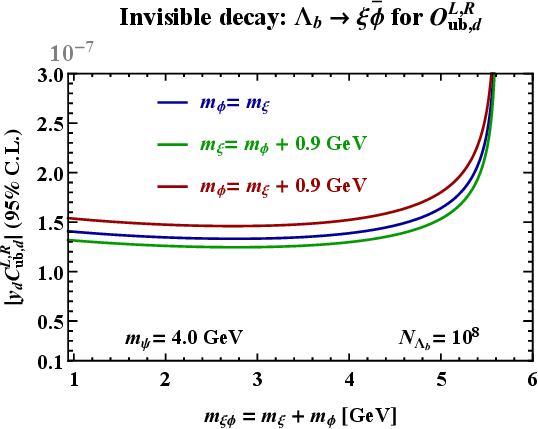}}
    \caption{ The constraints on the parameters $|y_d~\!C^{L,R}_{ud,b}|$ and $|y_d~\!C^{L,R}_{ub,d}|$ versus total mass $m_{\xi\phi}=m_\xi+m_\phi$ with $95\%$ C.L..
    As the DM stability prohibits a large mass splitting of dark fermion and dark baryons, we estimate the constraints in three different mass distribution of dark matters.
    }
    \label{fig:constraint-Wilson-coefficient-invisible}
\end{figure}

Assuming the production number of $\Lambda_b^0$ baryon using the double-tag method as $N_{\Lambda_b}=10^8$, the constraints on the parameters $|y_d~\!C^{L,R}_{ud_a,d_b}|$ can be given straightforwardly from Eq.~\eqref{eq_branching-ratio}, depending on the different masses of $\bar\psi$, $\bar\phi$ and $\xi$. 
The constraints 
with $95\%$ C.L. are displayed in FIG.~\ref{fig:constraint-mchi} and FIG.~\ref{fig:constraint-Wilson-coefficient-invisible}, for the variations of $m_{\psi}$ and $m_{\xi}+m_{\phi}$, respectively. 

In FIG.~\ref{fig:constraint-mchi}, the constraints of the parameters depending on $m_{\psi}$ are given for $|y_d~\!C_{ud,b}^{L,R}|$ in the left panel, and for $|y_d~\!C_{ub,d}^{L,R}|$ in the right panel. 
We estimate the constraints in six different mass distributions of dark matters. 
Three cases of masses of $\xi$ and $\phi$ are shown in the blue, green and red curves, with (i) $m_{\xi}=m_{\phi}=0.49 \,m_{\psi}$, (ii) $m_\xi=0.90\,m_\psi$, $m_\phi=0.08\,m_\psi$, and (iii) $m_\xi=0.08\,m_\psi$, $m_\phi=0.90\,m_\psi$.
Since the stable dark particles $\xi$ and $\bar{\phi}$ carry most of the mass of $\bar{\psi}$, the decay of proton into $\xi\bar{\phi}$ through an off-shell $\bar{\psi}$ particles is forbidden for $m_\psi\simeq 1$ GeV.
In addition, to directly illustrate the dependence of the constraints on the mass distribution of dark particles, other three cases of $m_\xi$ and $m_\phi$ are shown in orange, cyan and magenta curves, with (I) $m_\xi=m_\phi=0.25\,m_\psi$, (II) $m_\xi=0.4\,m_\psi,~\!m_\phi=0.1\,m_\psi$ and (III) $m_\xi=0.1\,m_\psi,~\!m_{\phi}=0.4\,m_\psi$.
For these cases, a truncation at small mass region is imposed to prevent the decay of proton into $\xi\bar{\phi}$.
Since the DM stability in Eq.~\eqref{equ_mass_splitting} prohibits a large mass difference between dark fermion $\xi$ and dark scalar baryon $\phi$, it also imposes a truncation at the large mass region with the specific mass distribution.
We found that the constraints on the parameters of B-Mesogenesis can reach $\mathcal{O}(10^{-7})$\;GeV$^{-2}$. 
Since the values of form factors arose from the left- and right-handed operators are the same except for a different sign, the constraints on the operators with different chirality are equal to each other. 
For the operators with different flavour structures, a factor of $2$ in the constraint on $|y_d~\!C^{L,R}_{ub,d}|$ arises from the half suppression of the decay constants $\lambda^{L,R}_{ub,d}$ in the comparison between Eqs.~\eqref{equ:InvisibleDecayConstant} and \eqref{equ:InvisibleDecayConstant2}. 
In the allowed mass region, we found that the differences of the constraints between various mass distribution are very small, indicating the fully invisible decay of $\Lambda_b$ baryon is insensitive to the mass distribution of dark particles.
The drop of the constraints at $m_\psi$ close to $m_{\Lambda_b}$ arises from a nearly on-shell transition of $\Lambda_b^0\to\bar{\psi}$, which significantly enlarge the invisible decay rate.

Similarly, the constraints of parameters  on the mass of $m_{\xi\phi}=m_\xi+m_\phi$ is shown in FIG.~\ref{fig:constraint-Wilson-coefficient-invisible}, for $|y_d~\!C^{L,R}_{ud,b}|$ and $|y_d~\!C^{L,R}_{ub,d}|$ in the left and right panels, respectively.
The three cases of (i) $m_\phi=m_\xi$, (ii) $m_\xi=m_\phi+0.9 \; \text{GeV}$, and (iii) $m_\phi=m_\xi+0.9 \; \text{GeV}$, are displayed in blue, green and red curves, respectively, with the mass of $\psi$ fixed as 4.0\;GeV. 
It is interesting that the curves are roughly flat at the region of $m_{\xi\phi}$ between 1\;GeV and 5\;GeV. 
Furthermore, the differences  between the three curves in each figure are also very small. 
These phenomenons indicate that the constraints on the parameters are insensitive to the mass distribution of $m_\xi$ and $m_\phi$. 
This is beneficial to search for the stable particles in B-Mesogenesis. 
If there is a deviation in the semi-invisible decays of bottom baryons or mesons, the fully invisible decay of $\Lambda_b^0$ can further explore the 
stable components which is inherent in B-Mesogenesis.
The declines of the constraints at the large mass region result from the suppression of phase space.

\section{Conclusion}
\label{sec:conclusion}

The similar densities of baryons and dark matter in the universe imply that they might arise from a same UV complete model. 
The B-Mesogensis, which assume dark matter can be charged under the baryon number, attempts to simultaneously explain the origins of dark matter and baryon asymmetry. 
Most of previous studies mainly concentrate on exploring the semi-invisible decays of $B$ mesons, since this model was proposed.
Actually, the invisible and semi-invisible decays of bottom baryons are also crucial for exploring the B-Mesogenesis at the GeV scale. 
In particular, as the B-Mesogenesis must involves a dark Majorana fermion and a dark scalar baryon to keep the DM and baryon asymmetry stable, the fully invisible decay of $\Lambda_b^0$ yield a distinctive signal to directly probe the stable particles of this scenario. 
What's more, when exploring the B-Mesogenesis in the decays of bottom baryons, the hadronic matrix elements are inevitable and vital for determining the physical observables. 
We systematically study the hadronic semi-invisible decays of $\Lambda_b^0\to \pi^0(K^0)\bar\psi$ and $\Xi_b^-\to \pi^-(K^-)\bar\psi$, the radiative semi-invisible decay of $\Lambda_b^0\to \gamma\bar\psi$, and the fully invisible decay of $\Lambda_b^0\to \xi\bar\phi$. The relevant hadronic matrix elements are calculated within the QCD factorization method in the heavy quark limit. 
Finally, we analyze the sensitivities on the parameters of B-Mesogenesis by searching for the invisible and semi-invisible decay of bottom baryons at future lepton colliders.
Assuming the production of bottom baryons by the double-tag method $N_{\Lambda_b,\Xi_b}=10^8$, we found that the constraints on the Wilson coefficients $|C^{L}_{ub,d}|$ and $|C^{L}_{ub,s}|$ in semi-invisible decay of bottom baryons can reach $\mathcal{O}(10^{-8})$ GeV$^{-2}$, while the constraints on $|y_d~\!C^{L,R}_{ud,b}|$ and $|y_d~\!C^{L,R}_{ub,d}|$ can reach $\mathcal{O}(10^{-7})$ GeV$^{-2}$ from the fully invisible decay of $\Lambda_b^0$. 
Although the invisible decay of $\Lambda_b^0$ baryon sensitive to more new physics parameters, we want to emphasize that this decay mode provides a distinctive signal to directly explore the stable dark particles of B-Mesogenesis.

\section{Acknowledgments}
We are grateful to Sheng-Qi Zhang and Man-Qi Ruan for useful discussions. The work is partly supported by the National Natural Science Foundation of China with Grant No.12335003, 12275277 and the National Key Research and Development Program of China under Contract No.2020YFA0406400 and 2023YFA1606000.

\end{document}